\documentclass[12pt,a4paper]{article}
\usepackage{charter}
\usepackage{array} 
\usepackage{graphics,graphpap}
\usepackage{graphicx}
\usepackage{color}
\usepackage{graphicx}
\usepackage{dcolumn}
\usepackage{epstopdf}
\usepackage{amsmath}
\usepackage{changes}
\usepackage{amsfonts}
\usepackage{textcomp}
\usepackage{bbm}
\usepackage{setspace}
\usepackage{dsfont}
\usepackage{feynmp}
\usepackage{amsthm,amsmath,amssymb,xcolor,sectsty,latexsym,mathtools,hyperref,mathrsfs,slashed,bm,epsfig,bigints,amsfonts}
\usepackage[font=small,skip=25pt]{caption}
\usepackage{longtable}
\usepackage{url}
\hypersetup{
    colorlinks = true,
    linktocpage,
    linkcolor= blue,
    citecolor = blue,
    urlcolor = blue
}
\usepackage{tocloft}

\cftsetindents{section}{0em}{2em}
\cftsetindents{subsection}{0em}{2em}

\newcommand{\beq}{\begin{eqnarray}}
\newcommand{\eeq}{\end{eqnarray}}
\newcommand{\be}{\begin{equation}}
\newcommand{\ee}{\end{equation}}

\def\drawbox#1#2{\hrule height#2pt
        \hbox{\vrule width#2pt height#1pt \kern#1pt
              \vrule width#2pt}
              \hrule height#2pt}

\def\Asym#1#2{\vcenter{\vbox{\drawbox{#1}{#2}
              \kern-#2pt 
              \drawbox{#1}{#2}}}}

\def\simge{\mathrel{%
   \rlap{\raise 0.511ex \hbox{$>$}}{\lower 0.511ex \hbox{$\sim$}}}}

\def\simle{\mathrel{
   \rlap{\raise 0.511ex \hbox{$<$}}{\lower 0.511ex \hbox{$\sim$}}}}

\def\s#1{\setbox0=\hbox{$#1$}%
\rlap{\ifdim\wd0>.7em\kern.22\wd0\else\kern.1\wd0\fi /}#1}

\begin{document}

\begin{titlepage}
\title{\vspace*{-2.0cm}
\hfill {\small }\\[3mm]
\bf\Large Degenerate Fermion Dark Matter from a Broken $U(1)_{\rm B-L}$ Gauge Symmetry
\\[1mm] }

\author{
Gongjun Choi$^{1}$\thanks{{\color{magenta}gongjun.choi@gmail.com}},~~
Motoo Suzuki$^1$\thanks{{\color{magenta}m0t@icrr.u-tokyo.ac.jp}},~~~and~~
Tsutomu. T. Yanagida$^{1,2}$\thanks{{\color{magenta}tsutomu.tyanagida@ipmu.jp}}
\\ \\
$^1${\normalsize \it Tsung-Dao Lee 
Institute, Shanghai Jiao Tong University, Shanghai 200240, China,}\\
$^2${\normalsize \it Kavli IPMU (WPI), UTIAS, The University of Tokyo,}\\
{\normalsize \it 5-1-5 Kashiwanoha, Kashiwa, Chiba 277-8583, Japan}\\
}
\date{\today}
\maketitle
\thispagestyle{empty}

\begin{abstract}
\noindent
The extension of the Standard model by assuming $U(1)_{\rm B-L}$ gauge symmetry is very well-motivated since it naturally explains the presence of heavy right-handed neutrinos required to account for the small active neutrino masses via the seesaw mechanism and thermal leptogenesis. 
Traditionally, we introduce three right handed neutrinos to cancel the $[U(1)_{\rm B-L}]^3$ anomaly. However, it suffices to introduce two heavy right-handed neutrinos for these purposes and therefore we can replace one right-handed neutrino by new chiral fermions to cancel the $U(1)_{\rm B-L}$ gauge anomaly. Then, one of the chiral fermions can naturally play a role of a dark matter candidate. 
In this paper, we demonstrate how this framework produces a dark matter candidate which can address the so-called ``core-cusp problem". As one of the small scale problems that $\Lambda$CDM paradigm encounters, it may imply an important clue for a nature of dark matter. One of resolutions among many is hypothesizing that sub-keV fermion dark matter halos in dwarf spheroidal galaxies are in (quasi) degenerate configuration. We show how the degenerate sub-keV fermion dark matter candidate can be non-thermally originated in our model and thus can be consistent with Lyman-$\alpha$ forest observation. Thereby, the small neutrino mass, baryon asymmetry, and the sub-keV dark matter become consequences of the broken B-L gauge symmetry.
\end{abstract}

\end{titlepage}

\tableofcontents
\newpage

\section{\label{sec:intro}Introduction}
Despite various evidences for the presence of dark matter (DM), DM's nature has not been uncovered yet. The central questions in regard to DM concern a mass of DM, non-gravitational interaction DM does, and its stability. Answers to these questions are considered essential factors in understanding not only a history and structure of the Universe in cosmology, but also a bigger and more fundamental picture lying behind the Standard model (SM) in particle physics. Seen from the perspective of this kind, DM related-observational anomalies reported in the study of cosmology and astrophysics could serve as a critical hint for physics beyond the Standard Model (BSM) although it is not necessary.

With that being said, a well known discrepancy between what has been expected based on a standard hypothesis of the cold dark matter (CDM) and what are observed regarding the small scale structure (galactic or sub-galactic scale) may deserve attention from a well-motivated BSM physics. Cuspy halo profiles of dwarf galaxies predicted by N-body simulations equipped with CDM~\cite{Navarro:1996gj,Fukushige:1996nr,Ishiyama:2011af} are at odds with the cored halo profiles implied by stellar kinematic data of low mass galaxies~\cite{Borriello:2000rv,Gilmore:2007fy,Oh:2008ww,deBlok:2009sp,Walker:2011zu}, which might be signaling a nature of DM deviating from collisionless and cold aspects.
Along with ``the missing satellite problem''~\cite{Moore:1999nt,Kim:2017iwr} and ``too-big-to-fail problem"~\cite{Boylan_Kolchin_2011}, this so-called ``core-cusp problem"~\cite{Moore:1999gc} is challenging for the most popular and robust CDM framework in spite of the success it has achieved thus far in accounting for the large scale structure (LSS) of the Universe and evolution thereof. 

Relying on predicted phenomenological consequences arising from a specific mass or non-gravitational interaction that DM enjoys, several alternative frameworks to CDM have been suggested so far in an effort to address the core-cusp problem (and other small-scale issues as well). These include warm dark matter (WDM)%
\footnote{Warm dark matter is assuming a DM particle characterized by a small enough mass to produce a free-streaming length of $\mathcal{O}(0.1)$Mpc and also by non-zero velocity dispersion. This feature enables WDM to erase density perturbations for the scale smaller than its free-streaming length, and thereby suppression of matter power spectrum on small scale and of formation of sub-halos is induced in comparison to CDM case~
\cite{Bode:2000gq,Colin:2000dn,AvilaReese:2000hg,Zavala:2009ms,Tikhonov:2009jq,Lovell:2011rd}
. As for the core size of a dwarf galaxy, WDM N-body simulations were conducted to study how the primordial velocity dispersion of WDM affects the inner structure of DM halo in ~\cite{Colin:2007bk,Maccio:2012qf}. In particular, sub-keV WDM is shown to produce the halo core size of $\mathcal{O}(100){\rm pc}$ for a typical sub-halo mass of Milky way whereas WDM with $1-2{\rm keV}$ mass does the core of 10-50pc~\cite{Maccio:2012qf}.}
\cite{Bode:2000gq,Colin:2000dn,Destri:2012yn}
, ultra-light bosonic DM~\cite{Ji:1994xh,Hu:2000ke,Hui:2016ltb}%
\footnote{The quantum pressure of the ultra-light bosonic dark matter supported by Heisenberg's uncertainty principle can help the self-gravitating system achieve stability against gravitational collapse.}and self-interacting DM (SIDM)%
\footnote{The self-interaction helps efficient heat conduction from the outer more energetic DM particles to the inner colder ones, which leads to redistribution of energy and angular momentum of DM particles. Consequently, as was shown in relevant simulations~\cite{Rocha:2012jg,Peter:2012jh}
, the central halo becomes less dense compared to CDM case and a cored halo profile forms accordingly.}
\cite{Spergel:1999mh}. 

Another interesting possibility of DM resulting in a cored halo profile in a low mass galaxy is the fermion DM in the quantum degenerate limit.
Along the similar line, the hypothesis of the fermion DM as the self-gravitating (quasi) degenerate gas was invoked in~\cite{Domcke:2014kla,Randall:2016bqw,Savchenko:2019qnn,DiPaolo:2017geq,Alam:2001sn}
to explain the kinematics of dwarf spheroidal galaxies (dSphs).
The fitting procedures for the kinematic data (stellar velocity dispersion and halo radius of dSphs) yielded sub-keV mass regime as the possible fermion DM mass (See also Refs.~\cite{Shao_2013,Alexander:2016glq,Giraud:2018gxl,Pal:2019tqq} for more studies about sub-keV fermion DM). Because of this, sub-keV fermion DM can serve as a class of solution to the core-cusp problem if it resides in dSphs nowadays with sufficiently low temperature so as to sit in the (quasi) degenerate state. This sub-keV mass regime encounters a severe constraint from the Lyman-$\alpha$ forest (see, e.g. \cite{Viel:2013apy,Baur:2015jsy,Irsic:2017ixq}), but the solution can still be viable provided that the sub-keV fermion DM is non-thermally originated and the free-streaming length of DM is not too large to be consistent with constraints derived from the Lyman-$\alpha$ flux power spectrum. 
The free-streaming length range $0.3{\rm Mpc}<\lambda_{\rm FS}<0.5{\rm Mpc}$ of DM  would be of interest since it can be consistent with non-vanishing matter power spectrum at large scales and avoid too many satellites of Milky way size halo \cite{Bringmann:2007ft,Cembranos:2005us,Colin:2000dn,Domcke:2014kla}.

Given the problem and one of the answers to it described above, the next question naturally thrown from the particle physics side could be probably whether a well-motivated extension of SM can accommodate such a sub-keV degenerate fermion WDM. In this work, 
we give our special attention to an extension of SM with a gauged $U(1)_{\rm B-L}$ symmetry. On top of SM particle contents, in its minimal form, the model contains two heavy right handed neutrinos ($\overline{N}_{i=1,2}$) and a complex scalar ($\Phi$) for breaking $U(1)_{\rm B-L}$. By means of this basic setting, the model is expected to accomplish the successful explanation for the small neutrino masses via seesaw mechanism~\cite{Yanagida:1979as,GellMann:1980vs,Minkowski:1977sc} and the baryon asymmetry via the thermal leptogenesis~\cite{Fukugita:1986hr}. Now for the purpose of making the theory anomaly-free and accommodating a non-thermal sub-keV fermion DM candidate, more chiral fermions are added to the model. 
The similar framework was studied in Refs.~\cite{Nakayama:2011dj,Nakayama:2018yvj,Choi:2020tqp} under the name of ``Number Theory Dark Matter''.

Beginning with the minimal set of new $U(1)_{\rm B-L}$ charged particle contents relative to SM, we shall search for all the possible sub-keV fermion DM production mechanisms and check consistency with Lyman-$\alpha$ forest observation. We gradually move to the next minimal scenario whenever an inconsistency is detected. Finally, we arrive at scenarios where multiple fundamental questions of small neutrino masses, baryon asymmetry and DM resolving small scale could be dealt with at once. We shall test consistency with Lyman-$\alpha$ forest data by computing free-streaming length of DM and constructing a map between thermal WDM mass and our sub-keV fermion DM mass. As an additional consistency check, we compute $\Delta N_{\rm eff}^{\rm BBN}$ contributed by the sub-keV fermion DM and ``would-be" temperature today of the DM candidate.

\section{Model}
\label{sec:model}
As the starting point of the task to extend the SM in a minimal way, we introduce a gauged $U(1)_{\rm B-L}$ symmetry. As the most elegant way of explaining the small neutrino masses, seesaw mechanism predicts the presence of heavy right-handed neutrinos~\cite{Yanagida:1979as,GellMann:1980vs,Minkowski:1977sc}. The advantage of extending the gauge symmetry group of SM by including the gauged $U(1)_{\rm B-L}$ lies in precisely this point. The theory can be naturally rendered gauge anomaly free when there exist three right-handed neutrinos with the opposite lepton number to that of active neutrinos. The other remarkable consequence that immediately follows here is that the presence of the heavy right-handed neutrinos can help us understand imbalance between baryon and anti-baryon abundance. Induced by the out-of-equilibrium decay of the right-handed neutrino, the primordial lepton asymmetry can be converted into baryon asymmetry by sphaleron transition~\cite{Fukugita:1986hr}.

Motivated by these attractive aspects, we consider a variant of SM with the gauge symmetry group ${\rm G}_{\rm gauge}={\rm SU(3)}_{c}\times {\rm SU(2)}_{L}\times {\rm U(1)}_{Y}\times {\rm U(1)}_{\rm B-L}$. Concerning the particle contents of the model, we begin with SM particle contents plus only two right-handed neutrinos, which is the most economical addition for the seesaw mechanism and the successful thermal leptogenesis~\cite{Frampton:2002qc}. In addition, we introduce one complex scalar to the model of which vacuum expectation value (VEV) causes the spontaneous breaking of $U(1)_{\rm B-L}$. Via Majorana Yukawa coupling, the complex scalar imposes masses to the two right handed neutrinos on $U(1)_{\rm B-L}$ breaking.
\begin{table}[ht]
\centering
\begin{tabular}{|c||c|c|c|c|c|c|c|c|c|} \hline
 & $L^{(1)}$ & $L^{(2)}$ & $L^{(3)}$ & $\overline{e}_{R}$ & $\overline{\mu}_{R}$ & $\overline{\tau}_{R}$ & $\overline{N}^{(1)}$ & $\overline{N}^{(2)}$ &$\Phi_{-2}$ \\
\hline
$Q_{\rm B-L}$      &  -1  &  -1  & -1 & +1 & +1 & +1  & +1 & +1 &-2\\
\hline
\end{tabular}
\caption{$U(1)_{\rm B-L}$ charge assignment to SM lepton $SU(2)_{L}$ doublets and singlets, two right-handed neutrino Weyl fields ($\overline{N}^{(i=1,2)}$) and the new complex scalar ($\Phi_{-2}$). The superscript and subscript are denoting the generation and $U(1)_{\rm B-L}$ charge, respectively.}
\label{table1} 
\end{table}
Of course, one is naturally tempted to make introduction of three right-handed neutrinos at the moment since it can satisfy the anomaly free condition of $U(1)_{\rm B-L}$ and simultaneously may be able to explain DM by taking the lightest right handed neutrino (sterile neutrino) as a candidate of the dark matter.%
\footnote{Indeed, this scheme has been discussed in many literatures (see, e.g.~\cite{deGouvea:2005er,Asaka:2005an,Kusenko:2010ik})} However, there is no natural and convincing reason for such a large mass disparity between the first two and the last right-handed neutrinos. 
Therefore, assuming only two right-handed neutrinos for the seesaw mechanism and the thermal leptogenesis, we need to find another way to accommodate DM candidate in the model.  For this purpose, recalling the necessity for making the model $U(1)_{\rm B-L}$ anomaly free is of a great help. What could help to render $U(1)_{\rm B-L}$ anomaly free on behalf of the third right-handed neutrino? Looking at the anomaly free condition for $U(1)_{\rm B-L}$ given in Eq.~(\ref{eq:anomalyfree}) below,\footnote{The first one is for cancellation of $U(1)_{\rm B-L}^{3}$ anomaly and the second one is for cancellation of gravitational $U(1)_{\rm B-L}\times[{\rm gravity}]^{2}$ anomaly. The anomalies of $U(1)_{\rm B-L}\times$[SM gauge interactions$]^2$ are canceled by the quark-sector contributions.}
\beq
\sum_{i} (Q_{{\rm B-L},i})^{3}=0\quad,\quad\sum_{i} Q_{{\rm B-L},i}=0\,,
\label{eq:anomalyfree}
\eeq
where the sum is over fermions charged under $U(1)_{\rm B-L}$, and referring to Table~\ref{table1}, one comes to realize that a new set of fermions to be added to the model should satisfy 
\beq
\sum_{i} (Q_{{\rm B-L},j})^{3}=+1\quad,\quad\sum_{i} Q_{{\rm B-L},j}=+1\,,
\label{eq:anomalyfree2}
\eeq
where the sum is over a new set of fermions. Here the solutions including vector-like fermions are out of our interest. 
Probing the cases to meet the two conditions in Eq.~(\ref{eq:anomalyfree2}) simultaneously leads us to the conclusion that minimum number of the new chiral fermions to be added is four.\footnote{Due to the Fermat's theorem, solutions with two additional Weyl fields do not exist. Solutions with three extra Weyl fields always contain two vector-like fermions.} In effect, the similar logic was studied in \cite{Nakayama:2011dj,Nakayama:2018yvj} and diverse combinations of possible $Q_{\rm B-L}$ values were found there. Given many options, for our work, we choose $Q_{\rm B-L}$ assignments shown in Table~\ref{table2} for the new chiral fermions.
\begin{table}[h]
\centering
\begin{tabular}{|c||c|c|c|c|} \hline
 & $\psi_{-9}$ & $\psi_{-5}$ & $\psi_{7}$ & $\psi_{8}$ \\
\hline
$Q_{\rm B-L}$      &  -9  &  -5  & +7 & +8\\
\hline
\end{tabular}
\caption{$U(1)_{\rm B-L}$ charge assignment to the new chiral fermions to be added to the model. The subscript is denoting $U(1)_{\rm B-L}$ charge assigned to each.}
\label{table2} 
\end{table}

Now the $U(1)_{\rm B-L}$ charge assignment given in Table~\ref{table1} and \ref{table2} leads to the following renormalizable Yukawa couplings between $\Phi_{-2}$ and fermions in the model charged under $U(1)_{\rm B-L}$
\be
\mathcal{L}_{\rm Yuk}=\sum_{i=2}^{3}\frac{1}{2}y_{N,ij}\Phi_{-2}\overline{N}^{(i)}\overline{N}^{(i)}+y_{1}\Phi_{-2}^{*}\psi_{-9}\psi_{7}+y_{2}\Phi_{-2}\psi_{-5}\psi_{7}+{\rm h.c.}\quad\,.
\label{eq:yukawa}
\ee
Once $U(1)_{\rm B-L}$ is spontaneously broken by acquisition of VEV ($<\!\!\Phi_{-2}\!\!>\equiv V_{\rm B-L}$) of $\Phi_{-2}$, there arise mass eigenstates $\psi_{7}, \chi$ and $\xi$ with
\be
\chi\equiv\frac{y_{1}}{\sqrt{y_{1}^{2}+y_{2}^{2}}}\psi_{-9}+\frac{y_{2}}{\sqrt{y_{1}^{2}+y_{2}^{2}}}\psi_{-5}\quad,\quad\xi\equiv\frac{-y_{2}}{\sqrt{y_{1}^{2}+y_{2}^{2}}}\psi_{-9}+\frac{y_{1}}{\sqrt{y_{1}^{2}+y_{2}^{2}}}\psi_{-5}\,.
\ee
Here $\chi$ and $\psi_{7}$ form a Dirac fermion $\Psi_{\rm H}\equiv(\chi,\psi_{7}^{*})^{\rm T}$ with a mass $m_{\Psi_{\rm H}}\simeq\sqrt{y_{1}^{2}+y_{2}^{2}}\,\,V_{\rm B-L}$. $\chi$ and $\xi$ are assigned {\it effective} $U(1)_{\rm B-L}$ charges
\be
Q_{\chi}=\frac{-5y_{2}^{2}-9y_{1}^{2}}{y_{1}^{2}+y_{2}^{2}}\quad,\quad Q_{\xi}=\frac{-5y_{1}^{2}-9y_{2}^{2}}{y_{1}^{2}+y_{2}^{2}}\,.
\label{eq:Qchixi}
\ee

Interestingly, $\psi_{8}$ is never mixed with other fermions in the model because $Q_{\rm B-L}$s of both $\psi_{8}$ and $\Phi_{-2}$ are even  while those of other fermions are odd. This makes $U(1)_{\rm B-L}$ broken down to the residual $Z_{2}^{\rm B-L}$ under which all fermions except for $\psi_{8}$ are odd. Thus, $\psi_8$ is perfectly stable. In accordance with this observation, we take $\psi_{8}$ as the DM candidate in our model. We attribute the stability of DM to even $Q_{\rm B-L}$ of $\psi_{8}$ from now on. $\psi_{8}$ obtains its mass via a higher dimensional operator 
\be
\frac{\kappa}{2} M_{P}\left(\frac{\Phi_{-2}}{M_{P}}\right)^{8}\psi_{8}\psi_{8}\quad\Longrightarrow\quad m_{\rm DM}=\kappa\times\left(\frac{V_{\rm B-L}}{10^{15}{\rm GeV}}\right)^{8}{\rm eV} \,,
\label{eq:DMmass1}
\ee
where $\kappa$ is a dimensionless coefficient. Here we see that $U(1)_{\rm B-L}$ breaking scale directly determine the DM mass.

To estimate the mass of $\xi$, we can try to figure out the smallest mass eigenvalue in terms of $V_{\rm B-L}/M_{P}$ by writing down 4$\times$4 mass matrix for the fermion field vector $\vec{F}\equiv (\psi_{-9},\psi_{-5},\psi_{7},\overline{N}^{(i)}$) formed by not only renormalizable, but higher dimensional operators consistent with assumed symmetries. Particularly, owing to the terms%
\footnote{We omit the dimensionless coefficients for the notational simplicity.}
\be
\label{eq:ximix}
\mathcal{L}=\Phi_{-2} \overline{N}^{(i)}\overline{N}^{(i)}+\frac{{\Phi_{-2}^{*}}^2}{M_{P}}\psi_{-5}\overline{N}^{(i)}\ ,
\ee
the mass of $\xi$ is given by
\be
m_{\xi}\simeq8.7\times10^{7}\times\left(\frac{V_{\rm B-L}}{10^{15}}\right)^{3}{\rm GeV}\,.
\label{eq:ximass}
\ee

Interestingly, we obtained the small mass of the dark matter in the sub-keV regime,
\be
m_{\rm DM}\simeq 0.3\,{\rm keV}\ ,
\ee
for $V_{\rm B-L}\simeq 2\times 10^{15}\,{\rm GeV}$ owing to its $B-L$ charge. Then, $m_{\xi}$ becomes as large as $\sim\mathcal{O}(10^{9}){\rm GeV}$.

\section{Sub-keV Fermion Dark Matter Production from Scattering of SM Particles?: Maybe Not}
\label{sec:SMscattering}
In the previous section, we discussed a well-motivated minimal extension of SM in which sub-keV fermion DM may arise. As was pointed out in the introduction, we go through the procedure to check cosmological history and free-streaming length of DM candidate in the model in order to see whether the minimal model is good enough to be consistent with cosmological constraints including Lyman-$\alpha$ forest data. If not, we would gradually move to a next-to-minimal model by enlarging particle contents. From this place, since we are interested in sub-keV DM mass regime, we assume $V_{\rm B-L}\simeq(2-3)\times10^{15}{\rm GeV}$ based on Eq.~(\ref{eq:DMmass1}),  which is also consistent with the observed neutrino masses.

\begin{figure}[h]
\centering
\vspace*{-30mm}
\includegraphics[width=0.8\textwidth]{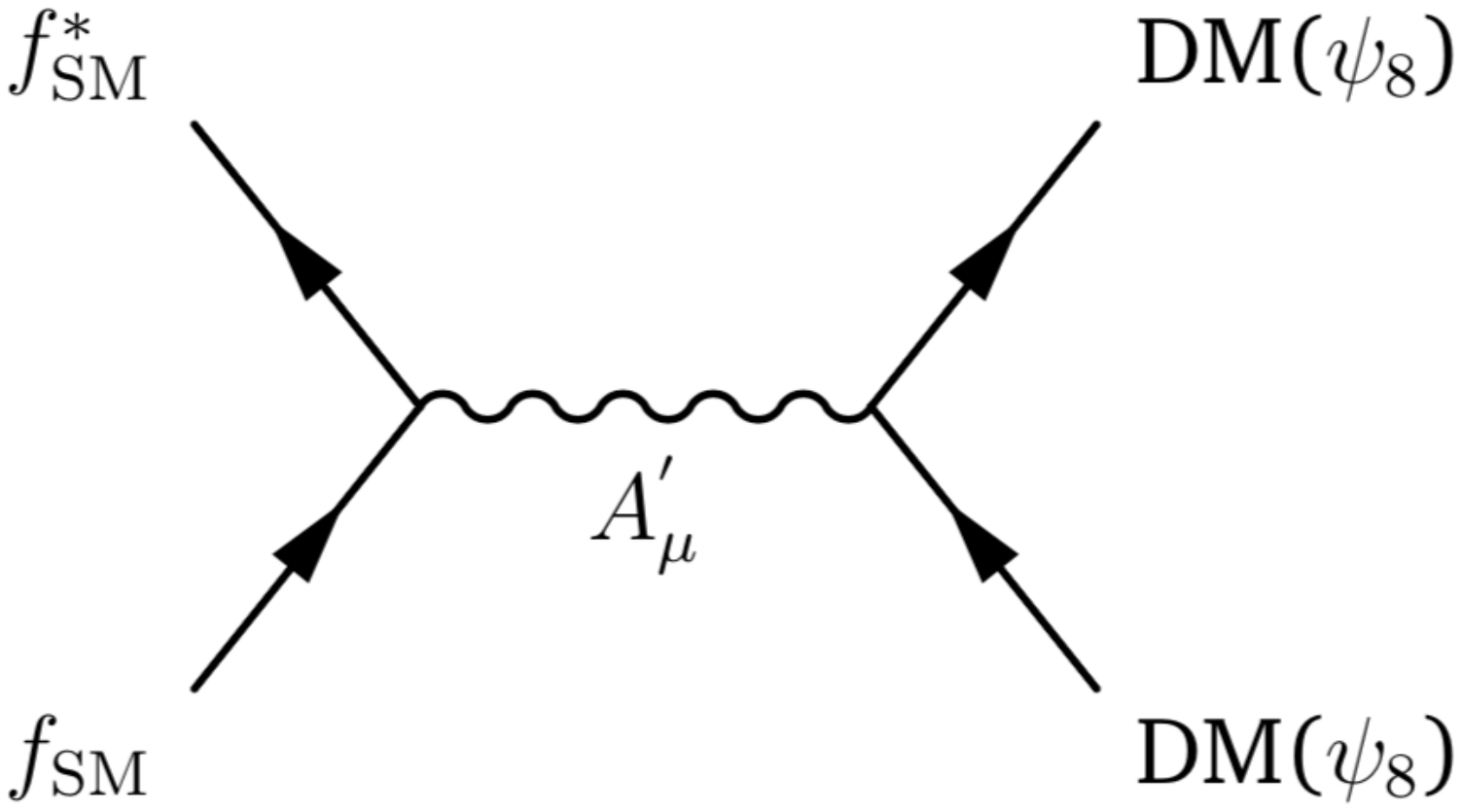}
\vspace*{-25mm}
\caption{DM ($\psi_{8}$) production from s-channel SM fermion scattering via $U(1)_{\rm B-L}$ gauge boson $A_{\mu}^{'}$ exchange.}
\vspace*{-1.5mm}
\label{fig1}
\end{figure}

In the minimal model, the only particle that is communicating with $\psi_{8}$ at the renormalizable level is $U(1)_{\rm B-L}$ gauge boson ($A_{\mu}^{'}$). Thus, the only way for $\psi_{8}$ to be produced is pair production resulting from scattering among the SM particles via the virtual $U(1)_{\rm B-L}$ gauge boson exchange.\footnote{This way of non-thermal production of DM pair is similar to the production mechanism of DM discussed in Refs.~\cite{Khalil:2008kp,Kusenko:2010ik,Nakayama:2011dj}. Particularly for the relevant Boltzmann equation solution, we refer the readers to \cite{Khalil:2008kp}.} The corresponding Feynman diagram is shown in Fig~\ref{fig1}. This production, however, should proceed with $\Gamma({\rm SM+SM}\rightarrow\psi_{8}+\psi_{8})<H$ at the reheating era. Otherwise, $\psi_{8}$ is thermalized by the SM thermal bath to become the thermal WDM which cannot have sub-keV mass regime. Thus we require
\be
\Gamma\simeq \frac{T_{\rm RH}^{5}}{V_{\rm B-L}^{4}}\lesssim\frac{T_{\rm RH}^{2}}{M_{P}}\simeq H\quad\Rightarrow\quad T_{\rm RH}\lesssim\left(\frac{V_{\rm B-L}^{4}}{M_{P}}\right)^{1/3}\simeq10^{14}{\rm GeV}\,.
\ee

For this production route which is most efficient at the reheating era, with the assumption that $\psi_{8}$ is identified as the sole DM component today, the DM number density to entropy density ratio reads~\cite{Kusenko:2010ik}
\be
Y_{\rm DM}\equiv\frac{n_{\rm DM}}{s_{\rm SM}}\sim\left.\frac{n_{f_{\rm SM}}\Gamma/H}{s_{\rm SM}}\right\vert_{T=T_{\rm RH}}\sim6.3\times10^{-7}\left(\frac{g_{*}}{100}\right)^{-3/2}\left(\frac{V_{\rm B-L}}{3\times10^{15}{\rm GeV}}\right)^{-4}\left(\frac{T_{\rm RH}}{10^{13}{\rm GeV}}\right)^{3}\,,
\label{eq:YDM}
\ee
where $n_{f_{\rm SM}}\sim T^{3}$ is the SM fermion number density, $\Gamma\equiv n_{f_{\rm SM}}<\!\!\sigma v\!\!>$ is the interaction rate for scattering among SM fermions, $g_{*}$ is the effective number of relativistic degrees of freedom and $s_{\rm SM}=2\pi^{2}g_{*}T^{3}/45$ is the entropy density. We assume the mass of $A_{\mu}^{'}$ is greater than a reheating temperature so that $A_{\mu}^{'}$ is never present in the SM thermal bath. Now the use of Eq.~(\ref{eq:YDM}) and $Y_{\rm DM}\equiv n_{\rm DM}/s_{\rm SM}\simeq4.07\times10^{-4}\times(m_{\rm DM}/1{\rm keV})^{-1}$ along with Eq.~(\ref{eq:DMmass1}) above yields%
\footnote{From $\Omega_{\rm{DM},0}=0.24$, $H_{0}=70\rm{km/sec/Mpc}$ and $s_{\rm{SM},0}\simeq2.945\times10^{-11}\rm{eV}^{3}$ (entropy density today), DM abundance $Y_{\rm{DM}}\equiv n_{\rm{DM}}/s_{\rm{SM}}$ is expressed in terms of DM mass as
\begin{equation}
Y_{\rm{DM}}\equiv\frac{n_{\rm{DM}}}{s_{\rm{SM}}}\simeq4.07\times10^{-4}\times\left(\frac{m_{{\rm DM}}}{1{\rm keV}}\right)^{-1}\,,
\label{eq:DMA1}
\end{equation}
at DM production time.}
\be
3.9\times10^{-3}\simeq\left(\frac{g_{*}}{100}\right)^{3/2}\left(\frac{T_{\rm RH}}{10^{13}{\rm GeV}}\right)^{-3}\left(\frac{m_{\rm DM}}{1{\rm keV}}\right)^{-1/2}\frac{1}{\sqrt{\kappa}}\,.
\ee
In other words, for given a $m_{\rm DM}$, $T_{\rm RH}$ required for production of the correct amount of DM abundance today via scattering among SM fermions must be
\be
T_{\rm RH}\simeq6.4\times10^{13}\times\left(\frac{g_{*}}{100}\right)^{1/2}\left(\frac{m_{\rm DM}}{1{\rm keV}}\right)^{-1/6}\left(\frac{1}{\kappa}\right)^{1/6}{\rm GeV}\,.
\label{eq:TRH1}
\ee
For $g_{*}\simeq100$, $m_{\rm DM}\simeq\mathcal{O}(100){\rm eV}$ and $\kappa\simeq\mathcal{O}(1)$, $T_{\rm RH}$ in Eq.~(\ref{eq:TRH1}) reads $\sim\mathcal{O}(10^{13}){\rm GeV}$. Provided that the right amount of DM is produced at the reheating era with this reheating temperature, we also expect production of $\xi$ via the similar SM scattering with the production ratio
\be
\frac{\Gamma({\rm SM+SM}\rightarrow\xi+\xi)}{\Gamma({\rm SM+SM}\rightarrow\psi_{8}+\psi_{8})}\simeq\frac{Q_{\xi}^{2}}{Q_{\psi_{8}}^{2}}=\left(\frac{-5y_{1}^{2}-9y_{2}^{2}}{8y_{1}^{2}+8y_{2}^{2}}\right)^{2}\,.
\label{eq:BR}
\ee
On production, we anticipate that $\xi$ completes decaying to SM Higgs and lepton before EW symmetry breaking time is reached and so it is cosmologically harmless (for detail, see Appendix~\ref{app:A}).
Due to the small interaction rate, $\psi_{8}$ starts free-streaming since production near the reheating era. The free-streaming length must be checked to be at least smaller than 0.5Mpc. This is for avoiding too much suppression of the matter power spectrum on small scales inconsistent with observation.

The free-streaming length of DM is computed by
\begin{eqnarray}
\lambda_{\rm FS}&=&\int_{t_{p}}^{t_{0}}\frac{<\!\!v_{\rm DM}(t)\!\!>}{a}{\rm d}t\cr\cr&\simeq&\int_{a_{p}}^{1}\frac{1}{H_{0}\sqrt{\Omega_{\rm rad,0}+a\Omega_{\rm m,0}}}\frac{<\!\!p_{\rm DM}(a_{p})\!\!>a_{p}}{\sqrt{(<\!\!p_{\rm DM}(a_{p})\!\!>a_{p})^{2}+m_{\rm DM}^{2}a^{2}}}{\rm d}a
\label{eq:FS1}
\end{eqnarray}
where $t_{p}$ and $a_{p}$ are the time and the scale factor at which DM starts free-streaming, $<\!\!v_{\rm DM}(t)\!\!>$ is the average velocity of the dark matter at the time $t$, and $<\!\!p_{\rm DM}(a_{p})\!\!>$ is the DM momentum at $t_p$. $\Omega_{\rm rad,0}$ and $\Omega_{\rm m,0}$ denote the radiation and matter density parameters, respectively.
Even if $\psi_{8}$ does not form a dark thermal bath, its momentum distribution is expected to be similar to the thermal distribution since it is produced from scattering of SM fermions which are in the thermal bath. 
The average momentum of the DM is estimated as
\be
<\!\!p_{\rm DM}(a_{p})\!\!>\,\,\gtrsim\,\,3.15\times T_{\rm RH}\,.
\label{eq:pdmatap}
\ee
where $a_{p}=a_{\rm RH}\simeq(10^{-13}\,{\rm GeV})/T_{\rm RH}$ is the time of the onset of DM free-streaming.\footnote{Here we use $a_{\rm EW}(a_{\rm EW})=a_{\rm RH}T_{\rm RH}$ with $a_{\rm EW}\simeq10^{-15}$ and $T(a_{\rm EW})\simeq100{\rm GeV}$.} The factor 3.15 applies for the typical thermal distribution of fermions. Using Eq.~(\ref{eq:pdmatap}), estimation of $\lambda_{\rm FS}$ for even $m_{\rm DM}=1$keV yields 1.25Mpc. The smaller DM mass corresponds to the longer $\lambda_{\rm FS}$ than this. This estimation concludes that the minimal scenario cannot produce a degenerate sub-keV fermion DM candidate for explaining the cored DM profiles for dSphs. 

Then, what another way could be considered to produce sub-keV fermion DM with a shorter free-streaming length? We notice that decreasing $T_{\rm RH}$ cannot shorten $\lambda_{\rm FS}$ in Eq.~(\ref{eq:FS1}) as long as $T_{\rm RH}\gtrsim10{\rm MeV}$ where 10MeV is the lower bound of $T_{\rm RH}$ from BBN. On the other hand, because $\psi_{8}$ cannot be coupled to any particle in the model other than $U(1)_{\rm B-L}$ gauge boson at the renormalizable level,\footnote{Note that $\psi_{8}$ cannot have a renormalizable coupling to a gauge singlet inflaton that satisfies gauge and Lorentz invariance all together.} no other DM production mechanism can be envisioned in the minimal model. Hence we cannot help but conclude that $\lambda_{\rm FS}$ cannot be shorten unless another DM production mechanism is considered by modifying the minimal model.

Therefore, we conclude that $\psi_{8}$ produced from SM particle scattering cannot be a candidate for the degenerate sub-keV DM to resolve the core-cusp problem. Now the whole of reasoning we followed in Sec.~\ref{sec:SMscattering} necessitates searching for a new way of producing DM which we discuss in the next section.

\section{Sub-keV Fermion DM from Inflaton Decay}
\label{sec:fermionDMinflaton}
As a next step, let us consider the DM production from the inflaton decay. $\psi_8$ can be coupled to the inflaton via
\begin{align}
\mathcal{L}\sim \frac{\Phi_I}{M_P}\psi_8^\dagger \bar\sigma^\mu D_\mu  \psi_8\,,
\end{align}
where $\Phi_{I}$ is the inflaton field and is assumed to be a gauge singlet and Lorentz scalar from here on. If the mass of the B-L gauge boson, $m_{B-L}$, is larger than the inflaton mass, $i.e.$ $m_{\rm B-L}>m_I$, the decay rate of the process ~$\Phi_I\to \psi_8+\psi_8^\dagger+\text{B-L charged particles (X)}$~ is
\begin{align}
\Gamma(\Phi_I\to \psi_8+\psi_8^\dagger+X)\sim \frac{m_{I}^7}{M_P^2V_{B-L}^4}\ .
\end{align}

\begin{figure}[t]
\centering
\hspace*{-5mm}
\includegraphics[width=0.7\textwidth]{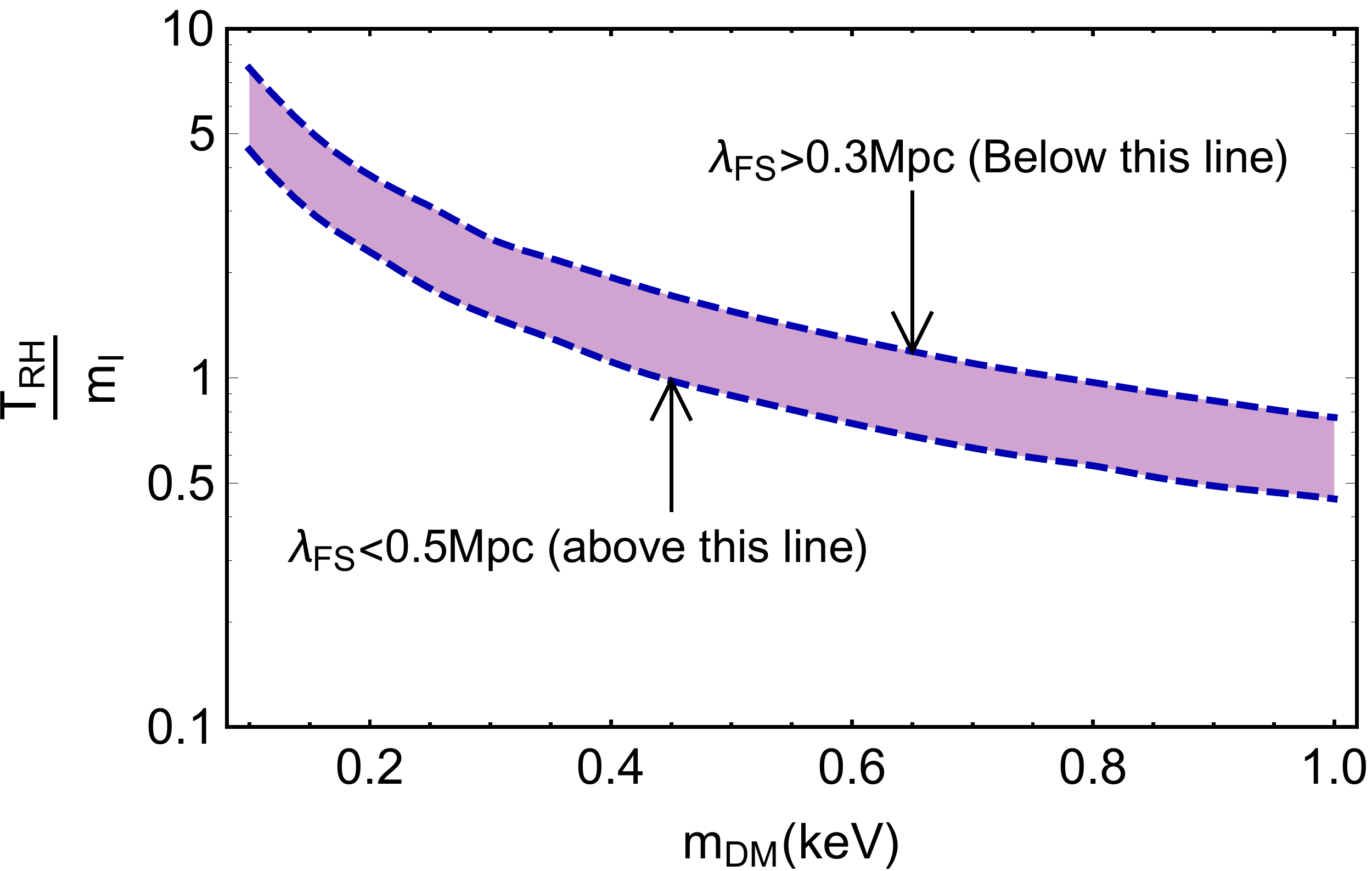}
\caption{The ratio of a reheating temperature ($T_{\rm RH}$) to an inflaton mass ($m_{I}$) that results in DM's free-streaming length $0.3{\rm Mpc}<\lambda_{\rm FS}<0.5$Mpc when DM is directly produced from the inflaton decay.}
\vspace*{-1.5mm}
\label{fig:rvsmDM}
\end{figure}

To explain the current abundance of the dark matter, one requires
\begin{align}
 \frac{m_{I}^7}{M_P^2V_{B-L}^4}\sim \frac{T_{\rm RH}^2}{M_P}{\rm Br}\sim \frac{T_{\rm RH} m_I}{M_P}10^{-4}\left(\frac{m_{\rm DM}}{1{\rm keV}}\right)^{-1}  \ ,
\end{align}
where ${\rm Br}$ is the branching ratio of $\Phi_I\to \psi_8+\psi_8^\dagger+X$ to the inflaton decay rate.%
\footnote{For the third relation, the branching ratio is determined to provide the current dark matter density (refer the discussion around Eq.\,(\ref{eq:DMA2})).} 
The reheating temperature is
\begin{align}
T_{\rm RH}\sim 10^4 m_I \frac{m_I^5}{M_P V_{\rm B-L}^4}\left(\frac{m_{\rm DM}}{1{\rm keV}}\right)\ .
\label{eq:reheating1}
\end{align}
To avoid the dominant production from the SM thermal bath (the case discussed in Sec.~\ref{sec:SMscattering}), $T_{\rm RH}\lesssim 10^{13}$\,GeV is required.
In addition, the ratio of a reheating temperature ($T_{\rm RH}$) to an inflaton mass ($m_{I}$) that results in DM's free-streaming length $0.3{\rm Mpc}<\lambda_{\rm FS}<0.5$Mpc is $\mathcal{O}(0.1)-\mathcal{O}(1)$ as can be seen in Fig.~\ref{fig:rvsmDM} when DM is directly produced from the inflaton decay with $<\!\!p_{\rm DM}(a_{p})\!\!>=m_{I}/2$ and $a_{p}=a_{\rm RH}$. 
However, this ratio with Eq.~(\ref{eq:reheating1}) for $T_{\rm RH}\lesssim 10^{13}$\,GeV leads to the condition
\be
T_{\rm RH}\ll m_I\ . 
\ee
This is inconsistent with Fig.~\ref{fig:rvsmDM}.
Thus, this possibility is out of our interest.

On the other hand, for $m_{\rm B-L}<m_I$, the decay rate is
\begin{align}
\Gamma(\Phi_I\to \psi_8+\psi_8^\dagger+A_\mu')\sim g_{\rm B-L}^2 m_I\left(\frac{m_I}{M_P}\right)^2\ .
\end{align}
where $g_{\rm B-L}$ denotes the B-L gauge coupling.
To explain the dark matter density, we require
\begin{align}
g_{\rm B-L}^2 m_I\left(\frac{m_I}{M_P}\right)^2\sim 10^{-4}\frac{T_{\rm RH} m_I}{M_P}\left(\frac{m_{\rm DM}}{1{\rm keV}}\right)^{-1}\ .
\end{align}
This leads to
\begin{align}
T_{\rm RH}\sim 10^4g_{\rm B-L}^2m_I\left(\frac{m_I}{M_P}\right)\left(\frac{m_{\rm DM}}{1{\rm keV}}\right)\ .
\label{eq:TRHMI2}
\end{align}
Similar to Eq.~(\ref{eq:reheating1}), Eq.~(\ref{eq:TRHMI2}) gives rise to
\begin{align}
T_{\rm RH}\ll m_I\ .
\label{eq:TRHMI3}
\end{align}
which is inconsistent with Fig.~\ref{fig:rvsmDM}. Thus, this possibility is also out of our interest.%
\footnote{Along with Eq.~(\ref{eq:TRHMI3}), too large a mass value itself for the inflaton also makes $\psi_{8}$ production from the inflaton decay with $m_{\rm B-L}<m_{I}$ not viable. From Fig.~\ref{fig:rvsmDM} and $m_{\rm B-L}<m_{I}$, we obtain $m_{\rm B-L}<m_{I}\sim(\mathcal{O}(0.1)-\mathcal{O}(1))T_{\rm RH}$ to have a degenerate fermion DM. If $A_{\mu}'$ produced from the process $\Phi_I\to \psi_8+\psi_8^\dagger+A_\mu'$ joins the SM thermal bath, $\psi_{8}$ would do so as well via the inverse decay process of $A_\mu'$ and becomes the thermal WDM. Thus we demand $\Gamma(A_\mu'\rightarrow\psi_8+\psi_8^\dagger)<H$ for $T_{\rm SM}\simeq m_{\rm B-L}$. In conjunction with Eq.~(\ref{eq:TRHMI2}) and the condition $m_{I}\sim(\mathcal{O}(0.1)-\mathcal{O}(1))T_{\rm RH}$, this requirement gives $m_{I}\sim\mathcal{O}(10^{15})-\mathcal{O}(10^{16}){\rm GeV}$ of which a corresponding inflation model is difficult to find.}

Therefore, we need to extend the minimal model to have the degenerate fermion DM.
As we will explain in detail, one simple possibility is to introduce a complex scalar field $\Phi_{16}$ with $Q_{\rm B-L}=16$.
This scalar field couples to DM through
\be
\mathcal{L}=y_*\Phi_{16}^*\psi_8\psi_8\, ,
\ee
where $y_*$ is a dimensionless coupling.%
\footnote{Similar to $\psi_{8}$, $\Phi_{16}$ could be produced from SM particle scattering as long as $T_{\rm RH}$ is large enough. The relevant diagram would be the one in Fig.~\ref{fig1} with $\psi_{8}$ replaced by $\Phi_{16}$. For this route, due to $Q_{\rm B-L}$ ratio, we expect four times more production of $\Phi_{16}$ than that of $\psi_{8}$. This case is also out of our interest because significant amount of DM ($\sim25\%$) would travel too large a free-streaming length as shown above using Eq.~(\ref{eq:FS1}) and (\ref{eq:pdmatap}).}

The renormalizable scalar sector potential we consider in the following reads\footnote{For the purpose of preventing $\Phi_{16}$ from being thermalized by any particle, we assume sufficiently suppressed renormalizable mixing of $\Phi_{16}$ with other scalars like $\sim(H^{\dagger}H)|\Phi_{16}|^{2}$ and $\sim|\Phi_{I}|^{2}|\Phi_{16}|^{2}$ which are allowed by symmetries in the model. See appendix.~\ref{app:B0} for more discussion about the Higgs portal couplings.} 
\be
V_{\rm scalar}=+m_{16}^{2}|\Phi_{16}|^{2}+\frac{\lambda_{16}}{4}|\Phi_{16}|^{4}+g\Phi_{I}|\Phi_{16}|^{2}+V(\Phi_{I})+V(H)\,,
\label{eq:Vscalar}
\ee
where $m_{16}$ is a parameter with a mass dimension, $\lambda_{16}$ is a dimensionless coupling, and $g$ is a parameter with a mass dimension, $V(\Phi_{I})$ and $V(H)$ are the potential for inflaton and SM Higgs doublet. We take  $<\!\!\Phi_{16}\!\!>=0$ in the vacuum, assuming the $\Phi_{16}$ has a positive mass squared, $m_{16}^{2} >0$. This makes Eq.~(\ref{eq:DMmass1}) intact. In Sec.~\ref{sec:fermionDMinflaton}, we assume that the Hubble induced mass squared for the $\Phi_{16}$ is positive so that $\Phi_{16}$ sits near the origin of the field space during and in the end of inflation.

Now $\Phi_{16}$ may be produced from the inflaton decay at the reheating era via the decay operator $\sim g\Phi_{I}|\Phi_{16}|^{2}$ if $m_{I}\gtrsim 2m_{16}$ holds. In this section, we attend to $\Phi_{16}$ particle produced in this manner. We are aiming to show that such a $\Phi_{16}$ could be a mother particle producing sub-keV fermion DM ($\psi_{8}$) consistent with Lyman-$\alpha$ forest observation. Depending on a value of $\lambda_{16}$, we have two different scenarios. We explore a case where a dark sector thermal bath forms in Sec.~\ref{sec:nonzerolambda16} and the other case where a dark sector thermal bath {\it never} forms in Sec.~\ref{sec:zerolambda16}.

\subsection{The Case with Formation of Dark Sector Thermal Bath}
\label{sec:nonzerolambda16}
In this section, we consider the case in which a dark thermal bath purely made up of $\Phi_{16}$ forms when $\Phi_{16}$ is produced from the inflaton decay. When $\lambda_{16}\neq0$ holds, from the comparison
\be
\Gamma\simeq\lambda_{16}^{2}T_{D}\gtrsim\frac{T_{\rm SM}^{2}}{M_{P}}\simeq H\quad\Rightarrow\quad x\lambda_{16}^{2}M_{P}\gtrsim T_{\rm SM}\,,
\ee
where $T_D$ is the temperature in the dark sector, we realize that it is easy for a dark thermal bath made up of $\Phi_{16}$ to form as far as the quartic interaction ($\lambda_{16}$) of $\Phi_{16}$ is not too small. Here $x=T_{D}/T_{\rm SM}$ is a fraction of order $\mathcal{O}(0.1)$ to be determined by DM relic density matching.
 We define the branching ratio Br to satisfy $n_{16}={\rm Br}\times n_{I}\simeq{\rm Br}\times(\rho_{{\rm SM}}/m_{I})$\footnote{This relation $n_{16}\simeq{\rm Br}\times(\rho_{{\rm SM}}/m_{I})$ can be used to derive relation between $m_{I}$, $T_{\rm RH}$ and Br. Using the approximation $2n_{\phi_{16}}=n_{\rm{DM}}$ at production time, one obtains
\begin{equation}
Y_{\rm{DM}}\equiv\frac{n_{\rm{DM,0}}}{s_{\rm{SM,0}}}\simeq\left.\frac{2n_{\phi_{16}}}{s_{\rm{SM}}}\right\vert_{T=T_{\rm RH}}\simeq\left.2{\rm Br}\frac{\rho_{{\rm SM}}}{m_{I}s_{\rm{SM}}}\right\vert_{T=T_{\rm RH}}\,.
\label{eq:DMA3}
\end{equation}
Using Eq.~(\ref{eq:DMA1}) and (\ref{eq:DMA3}), one obtains
\begin{equation}
\rm{Br}\frac{T_{\rm{RH}}}{m_{I}}\simeq2.7\times10^{-4}\times\left(\frac{m_{{\rm DM}}}{1{\rm keV}}\right)^{-1}\,.
\label{eq:DMA2}
\end{equation}
} where $n_{16}$ and $n_{I}$ are the number density of $\Phi_{16}$ and inflaton ($\Phi_{I}$) respectively. We assume $\rho_{{\rm I}}\simeq\rho_{{\rm SM}}$ at the reheating  era. From the number density comparison, we obtain the relation between dark sector temperature and SM sector temperature
\be
T_{{\rm D}}(a_{\rm RH})\simeq5.2\times{\rm Br}^{1/3}\times\frac{T_{\rm RH}^{4/3}}{m_{I}^{1/3}}\simeq 0.34\times\left(\frac{m_{{\rm DM}}}{1{\rm keV}}\right)^{-1/3}\times T_{\rm RH} \,,
\label{eq:Trelation}
\ee
where we used $g_{{\rm D}}(a_{\rm RH})=2$ and $g_{{\rm SM}}(a_{\rm RH})=106.75$. The second equality is coming from Eq.~(\ref{eq:DMA2}). The ratio of $T_{\rm D}/T_{\rm SM}$ remains the same until $\Phi_{16}$ decays to a DM pair. We note that Br is lower bounded as $\text{Br}\gtrsim (2.7\times 10^{-4}(m_{\rm DM}/1{\rm keV})^{-1})^2$.
This constraint is derived from the condition that $\Phi_{16}$ never gets into the SM thermal bath by the decay and the inverse decay process of $\Phi_I\leftrightarrow\Phi_{16}+\Phi_{16}^*$, and the requirement of obtaining the correct DM density (see Eq.~(\ref{eq:DMA2})).%
\footnote{The condition is ${\rm Br}\times T_{\rm RH}^2/M_P\lesssim m_I^2/M_P$ where the process $\Phi_I\leftarrow\Phi_{16}+\Phi_{16}^*$ is ineffective until the inflaton becomes non-relativistic and disappears.}

\begin{figure}[h]
\centering
\vspace*{-20mm}
\includegraphics[width=0.8\textwidth]{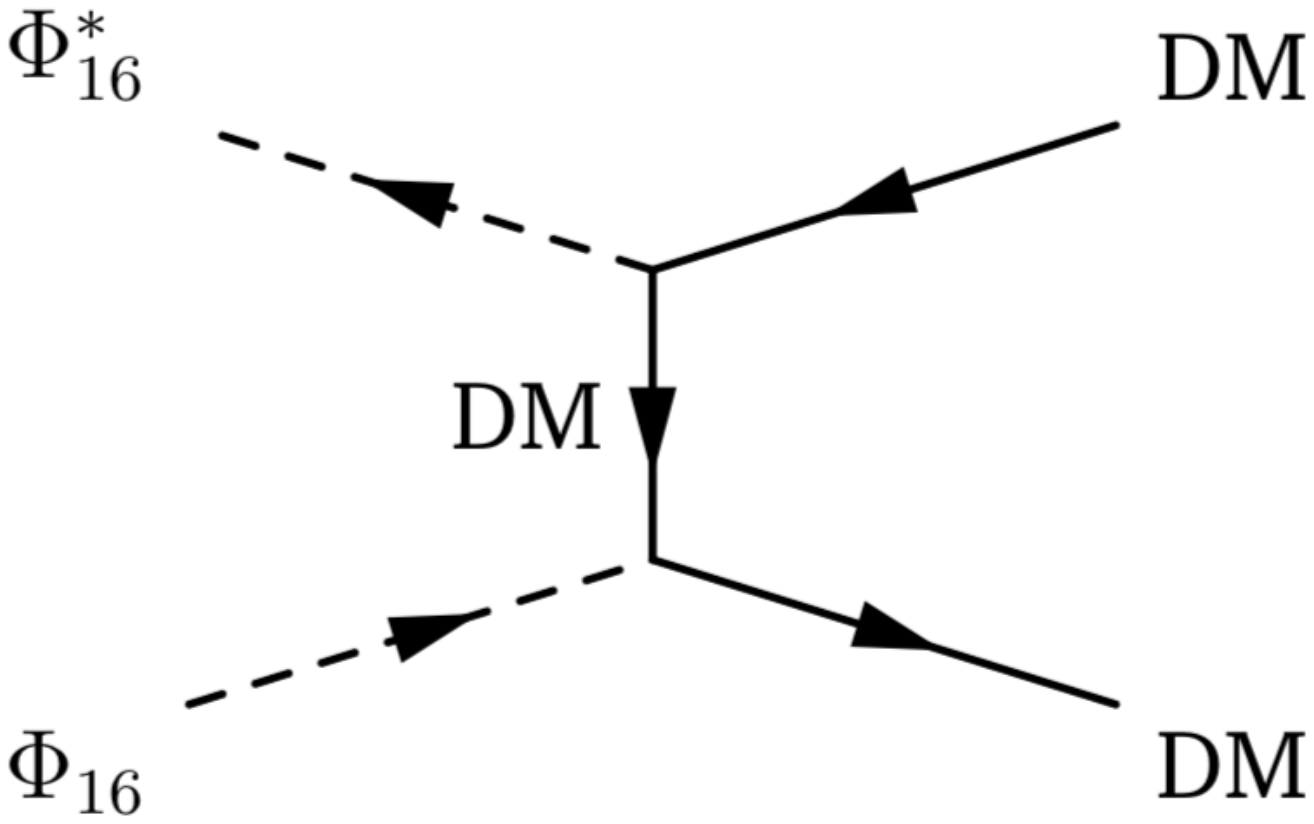}
\vspace*{-25mm}
\caption{The scattering among two $\Phi_{16}$s to produce a pair of DMs via t-channel DM exchange.}
\vspace*{-1.5mm}
\label{fig3}
\end{figure}

Concretely, we consider a scenario in which $\Phi_{16}$ becomes non-relativistic in the dark thermal bath before the time of $\Gamma(\Phi_{16}\rightarrow\psi_{8}+\psi_{8})\simeq H$ is reached. Afterwards, non-relativistic $\Phi_{16}$ decays to DM pair when the time of $\Gamma(\Phi_{16}\rightarrow\psi_{8}+\psi_{8})\simeq H$ is reached. The similar scenario was considered in \cite{Choi:2020tqp}. We demand that DM does not exist at the reheating era and is produced only from the decay of $\Phi_{16}$. To this end, define $T_{{\rm SM},i}$ ($T_{{\rm D},i}$) to be the SM (dark) thermal bath temperature at which $\Gamma_{i}\simeq H$ holds. For $\Phi_{16}+\Phi_{16}$ scattering to produce DM+DM and vice versa via t-channel DM exchange shown in Fig.~\ref{fig3}, the interaction rate reads,
\begin{equation}
\Gamma_{1}\simeq y_{*}^{4}T_{\rm D,1}\, ,\quad T_{{\rm SM},1}\simeq0.34\times\left(\frac{m_{{\rm DM}}}{1{\rm keV}}\right)^{-1/3}\times y_{*}^{4}M_{P}\, ,\quad T_{{\rm D},1}\simeq0.34^{2}\times\left(\frac{m_{{\rm DM}}}{1{\rm keV}}\right)^{-2/3}\times y_{*}^{4}M_{P}\,,
\label{eq:T1}
\end{equation}
where $y_{*}$ is Yukawa between $\Phi_{16}$ and DM. For $\Phi_{16}$ decay to DM+DM, the decay rate (when $m_{16}>T_{\rm DS}$) is given by
\begin{equation}
\Gamma_{2}\simeq\frac{y_{*}^{2}}{8\pi}m_{16}\, ,\quad T_{{\rm SM},2} \simeq\frac{y_{*}}{5}\sqrt{m_{16}M_{P}}\, ,\quad T_{{\rm D},2} \simeq0.34\times\left(\frac{m_{{\rm DM}}}{1{\rm keV}}\right)^{-1/3}\times\frac{y_{*}}{5}\sqrt{m_{16}M_{P}}\,, \label{eq:T2}   
\end{equation}
where $m_{16}$ is the mass of $\Phi_{16}$. 

To realize the scenario as we wish, we need to demand 
\be
m_{16}\,\,>\,\,T_{{\rm D},2}~~\text{and}~~m_{16}\,\,>\,\,T_{{\rm D},1}\,.
\label{eq:hierarchy1}
\ee
From the first inequality in Eq.~(\ref{eq:hierarchy1}), we obtain
\be
y_{*}\,\,<\,\,14.7\times\left(\frac{m_{{\rm DM}}}{1{\rm keV}}\right)^{1/3}\sqrt{\frac{m_{16}}{M_{P}}}\equiv y_{*,{\rm max}}\,.
\label{eq:firstinequality}
\ee
From the second inequality in Eq.~(\ref{eq:hierarchy1}), we obtain
\be
y_{*}\,\,<\,\,1.7\times\left(\frac{m_{{\rm DM}}}{1{\rm keV}}\right)^{1/6}\left(\frac{m_{16}}{M_{P}}\right)^{1/4}\,.
\label{eq:secondinequality}
\ee
In addition, requiring that DMs do not form a thermal bath via their self-interaction through $\Phi_{16}$ exchanges after its production
\begin{align}
\Gamma\simeq n_{\rm DM}\frac{y_*^4}{m_{16}^2}\,\,\lesssim\,\, H\,\,\text{at $a=a_{p}$}\ .
\end{align}
leads to the condition
\begin{align}
y_*\,\,\lesssim\,\,\left(\frac{m_{16}}{M_P}\right)^{3/10}\left(\frac{m_{\rm DM}}{1{\rm keV}}\right)^{1/5}\ .
\label{eq:thirdinequality}
\end{align}
Thus, we can see that for $m_{16}\lesssim 10^{13}$\,GeV, Eq.~(\ref{eq:secondinequality}) and (\ref{eq:thirdinequality}) are satisfied as long as Eq.~(\ref{eq:firstinequality}) is so. Together with $m_{16}$, $y_{*}$ is treated as a free parameter as far as $y_{*}\lesssim y_{*,{\rm max}}$ is satisfied. The smaller $y_{*}$ becomes, the later time onset of the free-streaming of DM becomes. Given a fixed initial momentum $<\!\!p_{\rm DM}(a_{\rm FS})\!\!>\simeq m_{16}/2$, the larger $a_{\rm FS}$ implies a larger $<\!\!p_{\rm DM}(a>a_{\rm FS})\!\!>$ for a fixed scale factor $a>a_{\rm FS}$. In the light of the fact that the late universe contribution to $\lambda_{\rm FS}$ is greater than the earlier one, we are led to speculate that for the same $(m_{16},m_{\rm DM})$, the smaller $y_{*}$ would lead to the larger $\lambda_{\rm FS}$ and hence more stringent constraint on $m_{\rm DM}$.

To constrain the model, we consider the free-streaming length criterion $0.3{\rm Mpc}<\lambda_{\rm FS}<0.5{\rm Mpc}$. Following Eq.~(\ref{eq:FS1}), the free-streaming length of DM produced from a non-relativistic $\Phi_{16}$ is
\begin{eqnarray}
\lambda_{\rm FS}&\simeq&\int_{t_{p}}^{t_{0}}\frac{<\!\!v_{\rm DM}(t)\!\!>}{a}{\rm d}t\cr\cr
&\simeq&\int_{a_{p}}^{1}\frac{1}{H_{0}\sqrt{\Omega_{\rm rad,0}+a\Omega_{\rm m,0}}}\frac{<\!\!p_{\rm DM}(a_{p})\!\!>a_{p}}{\sqrt{(<\!\!p_{\rm DM}(a_{p})\!\!>a_{p})^{2}+m_{\rm DM}^{2}a^{2}}}{\rm d}a\cr\cr&=&\int_{a_{p}}^{1}\frac{1}{H_{0}\sqrt{\Omega_{\rm rad,0}+a\Omega_{\rm m,0}}}\frac{m_{16}a_{p}}{\sqrt{(m_{16}a_{p})^{2}+4m_{\rm DM}^{2}a^{2}}}{\rm d}a\,,
\label{eq:FS2}
\end{eqnarray}
where $<\!\!p_{\rm DM}(a_{\rm FS})\!\!>\simeq m_{16}/2$ was used with $a_{p}\simeq a_{\rm FS}$. Here $a_{p}$ and $a_{\rm FS}$ are the scale factor at which the production of DM and the free-streaming of DM take place, respectively. Using Eq.~(\ref{eq:T2}), $a_{p}$ can be computed via
\be
a_{p}\simeq a_{\rm FS}\simeq\frac{10^{-13}{\rm GeV}}{T_{{\rm SM},2}}=\frac{5\times10^{-13}{\rm GeV}}{y_{*}\sqrt{m_{16}M_{P}}}\,.
\label{eq:aFS}
\ee

\begin{figure}[t]
\centering
\hspace*{-5mm}
\includegraphics[width=0.7\textwidth]{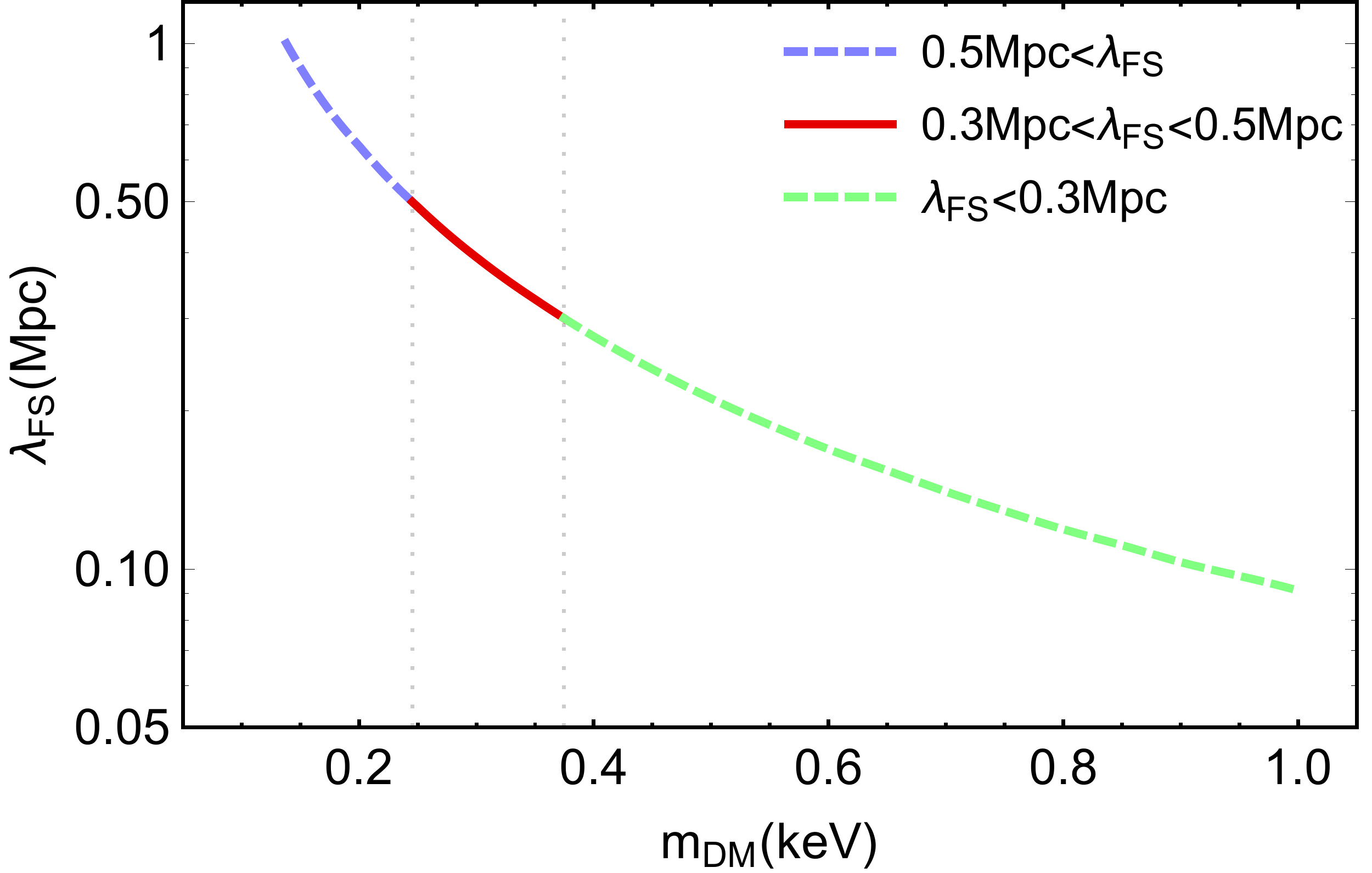}
\caption{Dark matter mass ($m_{\rm DM}$) vs Free streaming length ($\lambda_{\rm FS}$). For this plot, $y_{*}=y_{*,{\rm max}}$ in Eq.~(\ref{eq:firstinequality}) is assumed. For each $m_{\rm DM}$, the smaller $y_{*}$ yields the larger $\lambda_{\rm FS}$. }
\vspace*{-1.5mm}
\label{fig5}
\end{figure}

For a fixed ($m_{16},m_{\rm DM}$), $y_{*{\rm max}}$ in Eq.~(\ref{eq:secondinequality}) is determined, defining an allowed range of $y_{*}<y_{*{\rm max}}$. Within the range, the smaller $y_{*}$ results in the longer $\lambda_{\rm FS}$ since the free-streaming is delayed with the same initial momentum $<\!\!p_{\rm DM}(a_{\rm FS})\!\!>\simeq m_{16}/2$. This means that for each set of ($m_{16},m_{\rm DM}$), $y_{*}=y_{*,\rm max}$ in Eq.~(\ref{eq:aFS}) yields the smallest $\lambda_{\rm FS}$ value. On the other hand, for $y_{*}=y_{*,\rm max}$, we notice that $\lambda_{\rm FS}$ in Eq.~(\ref{eq:FS2}) becomes independent of $m_{16}$ since $m_{16}a_{p}$ is so. Thus, we realize that for $y_{*}=y_{*,\rm max}$, $\lambda_{\rm FS}$ is minimized for each $m_{\rm DM}$ whatever $m_{16}$ is. In Fig.~\ref{fig5}, we show $\lambda_{\rm FS}$ computed with $y_{*}=y_{*,{\rm max}}$ for the dark matter mass range $0.1{\rm keV}\lesssim m_{\rm DM}\lesssim1$keV. For a smaller $y_{*}$ choice, the curve in Fig.~\ref{fig5} would move upward. Without going through the further study with $y_{*}$ smaller than $y_{*,{\rm max}}$, we restrict ourselves to the case with $y_{*}=y_{*,{\rm max}}$ as an example, but the logic presented below can be also applied to other values of $(y_{*},m_{16})$ for the consistency check.

Starting with the momentum $\simeq m_{16}/2$ at $a=a_{\rm FS}$, the sub-keV DM we discuss here is still relativistic at BBN era with the momentum $\sim\mathcal{O}(1){\rm MeV}$. As such, the sub-keV DM serves as an extra-radiation during BBN era and therefore its contribution to $\Delta N_{\rm eff}^{\rm BBN}$ needs to be checked to be consistent with the known constraint. For each $m_{\rm DM}$, we computed $\Delta N_{\rm eff}^{\rm BBN}$ contributed by DM at the BBN era and found that the model with $y_{*}=y_{*,{\rm max}}$ is consistent with $\Delta N_{\rm eff}^{\rm BBN}\lesssim0.114$ (95\% C.L.) recently reported in \cite{Fields:2019pfx}. For computation of $\Delta N_{\rm eff}^{\rm BBN}$ contributed by DM, we refer the readers to Appendix \ref{app:B}. As the final consistency check, we estimated the ``would-be" temperature today ($\tilde{T}_{{\rm DM,0}}$) for $\psi_{8}$ based on Eq.~(\ref{eq:wouldbeT}) which reads
\be
\tilde{T}_{{\rm DM,0}}\simeq3.4\times10^{-9}\times\left(\frac{m_{\rm DM}}{1{\rm keV}}\right)^{-5/3}\,\,{\rm K}\,,
\ee
where we used $y_{*}=y_{*,{\rm max}}$ in Eq.~(\ref{eq:firstinequality}) and $a_{\rm FS}$ in Eq.~(\ref{eq:aFS}). We presented a brief explanation as to the necessary condition for fermion DM to be in degenerate configuration in Appendix \ref{app:D}. For $m_{\rm DM}$ of our interest, we see that $\tilde{T}_{{\rm DM,0}}<T_{{\rm DEG}}\simeq\mathcal{O}(10^{-4}){\rm K}-\mathcal{O}(10^{-3}){\rm K}$. This confirms that the current temperature of DM becomes low enough to accomplish degenerate configuration when structure formation is ignored.

We notice that $y_{*}$ can be constrained by $\Delta N_{\rm eff}^{\rm BBN}$, which we do not explore in detail. Intriguingly, for $y_{*}=y_{*,{\rm max}}$, the criterion $0.3{\rm Mpc}\lesssim\lambda_{\rm FS}\lesssim0.5{\rm Mpc}$ gives the mass constraint $0.25{\rm keV}\lesssim m_{\rm DM}\lesssim0.37{\rm keV}$, which lies in the range of degenerate fermion DM mass accounting for the cored DM profiles of dSphs in Refs.~\cite{Domcke:2014kla,Randall:2016bqw,Savchenko:2019qnn,DiPaolo:2017geq}. Another choice of $y_{*}<y_{*,{\rm max}}$ will make $0.3{\rm Mpc}\lesssim\lambda_{\rm FS}\lesssim0.5{\rm Mpc}$ correspond to a larger $m_{\rm DM}$ range.

Additionally, we also discuss the constraint on the mass of our DM candidate ($\psi_{8}$) mapped from a conservative lower bound for the mass of the thermal WDM, i.e. $1.9{\rm keV}$ (95\% C.L.), recently reported in \cite{Garzilli:2019qki}. We make a detail discussion about how the mapping can be achieved in Appendix~\ref{app:C}. Here we directly construct the map based on Eq.~(\ref{eq:Tmapping1}). We begin by equating the warmness parameters for $\psi_{8}$ ($\sigma_{\psi_{8}}$) and the thermal WDM ($\sigma_{\rm wdm}$)
\begin{equation}
\sigma_{\psi_{8}}=\sigma_{\rm wdm} \Longleftrightarrow{} \tilde{\sigma}_{\psi_{8}}\frac{T_{_{\psi_{8}}}}{m_{\psi_{8}}}=\tilde{\sigma}_{\rm wdm}\frac{T_{\rm wdm}}{m_{\rm wdm}}\,,
\label{eq:Tmapping2}
\end{equation} 
where $m$ and $T$ denote a mass and temperature, and $\tilde{\sigma}$ is defined in Eq.~(\ref{eq:sigmatilde}). As a particle produced from the decay of a non-relativistic mother particle, $\psi_{8}$ is characterized by the momentum space distribution function $f(q,t)=(\beta/q){\rm exp}(-q^{2})$ where $\beta$ is a normalization factor and $q\equiv p/T$ is used \cite{Kaplinghat:2005sy,Strigari:2006jf,Aoyama:2011ba,Kamada:2013sh,Merle:2013wta}. This gives us $\tilde{\sigma}_{\psi_{8}}\simeq1$. On the other hand, since $m_{16}>>m_{\rm DM}$ is assumed, DM temperature at the matter-radiation equality can be written as
\be
T_{\psi_{8}}(a_{\rm eq})=\frac{m_{16}a_{\rm FS}}{2a_{\rm eq}}=0.17\times10^{-7}\,{\rm keV}\times\left(\frac{m_{\psi_{8}}}{1{\rm keV}}\right)^{-1/3}\times(1+z_{\rm eq})\,,
\label{eq:Tpsi8}
\ee
with $a_{\rm FS}$ defined in Eq.~(\ref{eq:aFS}). Finally, by using $\tilde{\sigma}_{\rm wdm}=3.6$ for the thermal WDM and $T_{\rm wdm}(a_{\rm eq})=T_{\rm wdm,0}/a_{\rm eq}$ in Eq.~(\ref{eq:Tmapping2}), we obtain the map
\be
m_{\psi_{8}}\simeq 0.2\times m_{\rm wdm}\, .
\ee
Applying the conservative constraint $m_{\rm wdm}>1.9{\rm keV}$ \cite{Garzilli:2019qki}, we obtain $m_{\rm \psi_{8}}\gtrsim 0.4{\rm keV}$. This result may seem a tension with $m_{\rm DM}$ required for a degenerate fermion DM in \cite{Domcke:2014kla,Randall:2016bqw}. However, indeed there exist some uncertainties in velocity anisotropy parameter used for fitting of the stellar velocity dispersion, the lower bound of Fornax dSphs halo radius and baryon's effect on the DM halo profile. Also still for some dSphs other than Fornax, the best fitting for the stellar velocity dispersion is done by $m_{\rm DM}$ as large as 550-650eV \cite{Savchenko:2019qnn}. Here without performing a detailed fitting analysis to infer the degenerate fermion DM mass, we take a conservative attitude to understand $100{\rm eV}\!\lesssim m_{\rm DM}\!\lesssim1{\rm keV}$ as the interesting range relating to degenerate fermion DM solution to the core-cusp problem.

\subsection{The Case without Formation of Dark Sector Thermal Bath}
\label{sec:zerolambda16}
For the case where $\Phi_{16}$ does have a tiny or vanishing quartic interaction, $\Phi_{16}$ would not form a dark thermal bath as far as Yukawa interaction with $\psi_{8}$ is sufficiently small. Since production from the inflaton decay, it would continue to free-stream until it decays to a pair of $\psi_{8}$. Note that this early free-streaming of $\Phi_{16}$ is not problematic at all for the small scale perturbations since the early time free-streaming length is negligibly small. With this picture in mind, in this section, we study the possibility of having degenerate fermion DM arising from the decay of a free non-relativistic scalar $\Phi_{16}$. We explore the parameter space of the model where the free-streaming length of $\psi_{8}$ becomes consistent with Lyman-$\alpha$ forest observation.

In order to avoid having the thermal WDM, we focus on the scenario where $\Phi_{16}$ starts the free-streaming once produced from the inflaton decay. After that $\Phi_{16}$ becomes non-relativistic first and then decays to DM pairs. Differing from the previous case with $\lambda_{16}\neq0$, the time when $\Phi_{16}$ becomes non-relativistic is sensitive to inflaton mass now. $\Phi_{16}$ has momentum $p_{16}(a_{\rm RH})\simeq m_{I}/2$ at the reheating era on production and then becomes non-relativistic at 
\be
a=a_{\rm NR}\equiv\frac{m_{I}a_{\rm RH}}{2m_{16}}\simeq\frac{m_{I}\times10^{-13}{\rm GeV}}{2\times T_{\rm RH}\times m_{16}}\simeq\frac{{\rm Br\times\left(\frac{m_{{\rm DM}}}{1{\rm keV}}\right)\times10^{-13}{\rm GeV}}}{5.4\times10^{-4}\times m_{16}}\,,
\label{eq:aNR}
\ee
where we used $a_{\rm RH}\simeq(10^{-13}{\rm GeV})/T_{\rm RH}$ for the third equality and Eq.~(\ref{eq:DMA2}) for the last equality. 
For our purpose, we demand that
\be
T_{\rm SM}(a_{\rm NR})\,\,>\,\,T_{\rm SM,2}\,\,>\,\,T_{\rm SM,1} 
\label{eq:hierarchy3}
\ee
where $T_{\rm SM,1}$ and $T_{\rm SM,2}$ were defined in Eq.~(\ref{eq:T1}) and Eq.~(\ref{eq:T2}). From the first inequality in Eq.~(\ref{eq:hierarchy3}), we obtain
\be
y_{*}\,\,<\,\,\frac{2.7\times10^{-3}}{{\rm Br}\times\left(\frac{m_{{\rm DM}}}{1{\rm keV}}\right)}\times\sqrt{\frac{m_{16}}{M_{P}}}\equiv y_{*,1}\,.
\label{eq:ystar1}
\ee
From the second inequality in Eq.~(\ref{eq:hierarchy3}), we obtain
\be
y_{*}\,\,<\,\,0.84\times\left(\frac{m_{{\rm DM}}}{1{\rm keV}}\right)^{1/9}\left(\frac{m_{16}}{M_{P}}\right)^{1/6}\equiv y_{*,2}\,.
\label{eq:ystar2}
\ee
In addition, as discussed in Sec.~\ref{sec:nonzerolambda16}, we require
\begin{align}
y_*\,\,\lesssim\,\,\left(\frac{m_{16}}{M_P}\right)^{3/10}\left(\frac{m_{\rm DM}}{1{\rm keV}}\right)^{1/5}\equiv y_{*,3}\ ,
\label{eq:ystar3}
\end{align}
so that the DM does not form a dark thermal bath via their self-interaction through $\Phi_{16}$ exchange after its production.

\begin{figure}[t]
\centering
\hspace*{-5mm}
\includegraphics[width=0.7\textwidth]{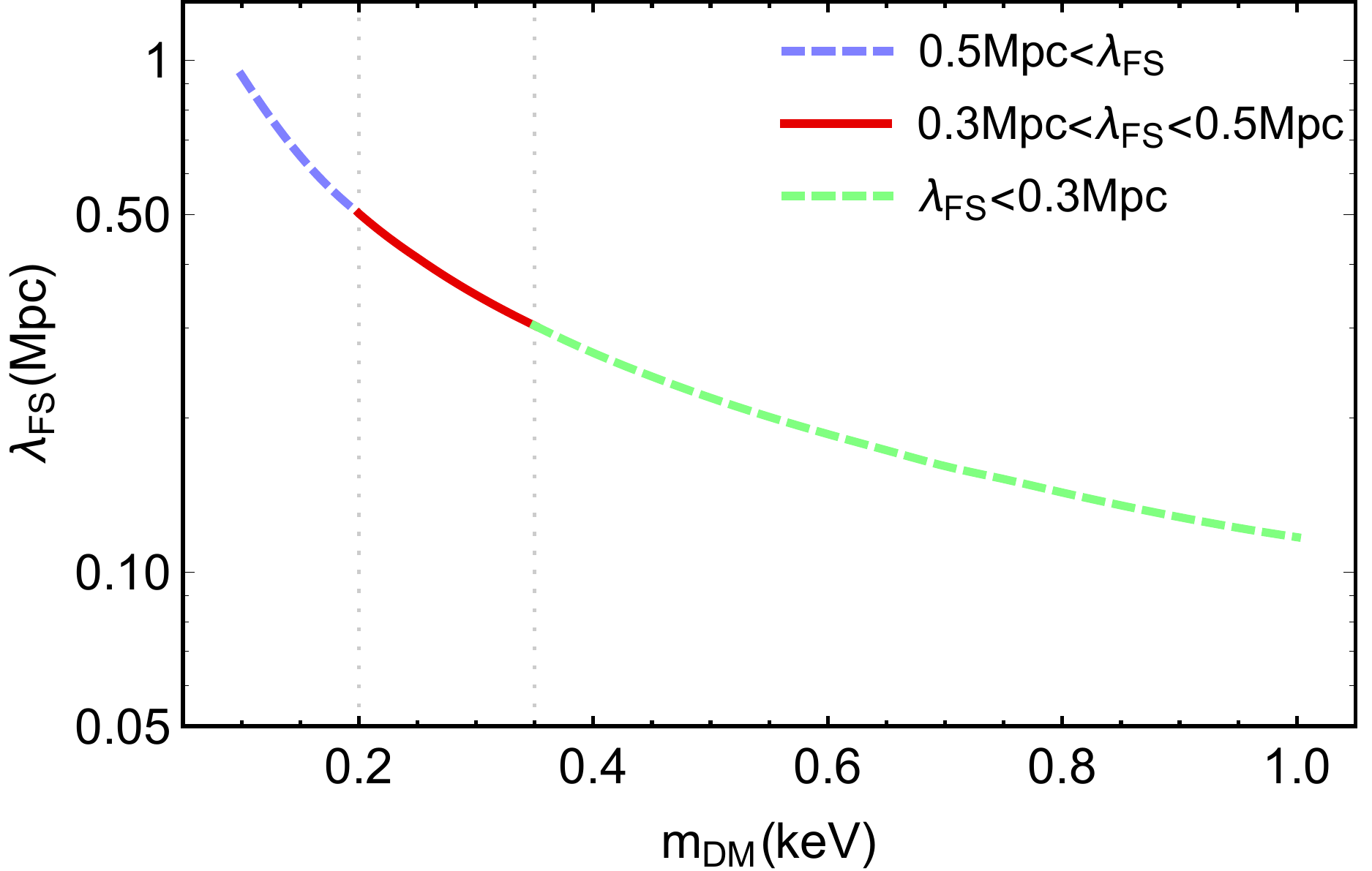}
\caption{Dark matter mass ($m_{\rm DM}$) vs Free streaming length ($\lambda_{\rm FS}$). For this plot, $(m_{16}\!=\!5\times10^{5}{\rm GeV},{\rm Br}\!=\!10^{-6},y_{*}\!=\!5\times10^{-6})$ is assumed. For each $m_{\rm DM}$, the smaller $y_{*}$ yields the larger $\lambda_{\rm FS}$.}
\vspace*{-1.5mm}
\label{fslvsmDM2}
\end{figure}

For a given set of $(m_{16},m_{\rm DM},{\rm Br})$, each of $(y_{*,1},y_{*,2},y_{*,3})$ is to be determined. Define $y_{*,{\rm max}}\equiv{\rm min}(y_{*,1},y_{*,2},y_{*,3})$. Then a choice of Yukawa coupling satisfying $y_{*}<y_{*,{\rm max}}$ will satisfy Eq.~(\ref{eq:hierarchy3}). Numerically we find that (1) for ${\rm Br}\gtrsim10^{-3}$, $y_{*,{\rm max}}=y_{*,1}$ for any sub-keV $m_{\rm DM}$ and (2) for ${\rm Br}\lesssim10^{-4}$, $y_{*,{\rm max}}$ is either $y_{*,1}$ or $y_{*,3}$. For a fixed $m_{\rm DM}$, $\lambda_{\rm FS}$ depends on $m_{16}$ and $y_{*}$, and these two are inversely-correlated. Thus, in principle, for a fixed $m_{\rm DM}$, a set of $(m_{16},y_{*})$ satisfying $\lambda_{\rm FS}\in(0.3,0.5)$Mpc can be readily found and consistent insofar as $y_{*}<y_{*,{\rm max}}$. In this section, instead of probing all the allowed parameter space for $(m_{16},{\rm Br},y_{*},m_{\rm DM})$, for our purpose it suffices to choose a specific benchmark set of parameters $(m_{16}\!=\!5\times10^{5}{\rm GeV},{\rm Br}\!=\!10^{-6},y_{*}\!=\!5\times10^{-6})$ to show that a degenerate sub-keV fermion DM can be produced in the model. Then we see that $y_{*}<y_{*,{\rm max}}$ is satisfied. We emphasize that this example is not atypical and the following logic and consistency check can also apply for other values of parameters. The result of computation for $\lambda_{\rm FS}(m_{\rm DM})$ is shown in Fig.~\ref{fslvsmDM2}. Interestingly, the range $0.2{\rm keV}\lesssim m_{\rm DM}\lesssim0.35{\rm keV}$ corresponds to the criterion $0.3{\rm Mpc}\lesssim\lambda_{\rm FS}\lesssim0.5{\rm Mpc}$ gives the mass constraint , which lies in the range of degenerate fermion DM mass accounting for the cored DM profiles of dSphs in Refs.~\cite{Domcke:2014kla,Randall:2016bqw,Savchenko:2019qnn,DiPaolo:2017geq}. The smaller $y_{*}$ and the larger $m_{16}$ would make the curve in Fig.~\ref{fslvsmDM2} move upward.

As the final consistency check, we compute $\Delta N_{\rm eff}^{\rm BBN}$ and ``would-be" temperature today for $\psi_{8}$. Firstly, from Eq.~(\ref{eq:aFS}), (\ref{eq:Neff}) and $y_{*}=5\times10^{-6}$, $\Delta N_{\rm eff}^{\rm BBN}$ is found to be at most $\simeq0.02$. This result is consistent with $\Delta N_{\rm eff}^{\rm BBN}\lesssim0.114$ (95\% C.L.) \cite{Fields:2019pfx}. Next, from Eq.~(\ref{eq:wouldbeT}), DM's ``would-be" temperature today reads $\tilde{T}_{{\rm DM,0}}\sim\mathcal{O}(10^{-9}){\rm K}-\mathcal{O}(10^{-8}){\rm K}$, which is smaller than $T_{{\rm DEG}}\simeq\mathcal{O}(10^{-4}){\rm K}-\mathcal{O}(10^{-3}){\rm K}$. Similarly to the case of Sec.~{\ref{sec:nonzerolambda16}}, this shows that the current temperature of DM becomes low enough to accomplish the degenerate configuration when structure formation is ignored.

\section{Sub-keV Fermion DM from Decay of a Scalar Field Coherent Oscillation}
\label{sec:fermionDMCO}
So far we have assumed that $\Phi_{16}$ has a positive Hubble induced mass squared during the inflation. However, we assume the negative Hubble induced mass squared in this section.
We consider the potential of $\Phi_{16}$,
\begin{align}
V=\left(m_{16}^2-c_2 H_{\rm inf}^2\right)|\Phi_{16}|^2+c_{2n} \frac{1}{(n!)^2}\frac{|\Phi_{16}|^{2n}}{M_P^{2n-4}}\ ,
\label{eq:V16}
\end{align}
where $H_{\rm inf}$ is the Hubble parameter during inflation, $n$ is a positive integer larger than one, $c_2$ and $c_{2n}$ are positive dimensionless couplings.
Then, $\Phi_{16}$ sits around the potential minimum with the amplitude $\Phi_{16,I}$ during the inflation,
\begin{align}
\Phi_{16,I}\simeq \left(\frac{(n!)^2c_2}{n\,c_{2n}}H_{\rm inf}^2M_P^{2n-4}\right)^{\frac{1}{2n-2}}\ ,
\end{align}
where we ignore the mass term with $m_{16}$ by assuming $m_{16}^2\ll c_2 H_{\rm inf}^2$.
After the end of inflation, the field value of $\Phi_{16}$ is given by
\begin{align}
\label{eq:Phi16amp}
\langle \Phi_{16}\rangle\simeq  \left(\frac{(n!)^2c_2}{n\,c_{2n}}H^2M_P^{2n-4}\right)^{\frac{1}{2n-2}}\ ,
\end{align}
for $n\geq 4$. 
Here, $H$ denotes the Hubble expansion rate.
This behavior of the scalar field is called the scaling solution~\cite{Liddle:1998xm,Harigaya:2015hha,Ibe:2019yew}. We focus on this scaling solution with $n=4$ as an example in the rest of this section.%
\footnote{We ignore the other terms with $n\neq 4$ not to affect the dynamics of $\Phi_{16}$. The analysis for the potential with $n=2~\text{or}~3$ will be given elsewhere.}
As the Hubble expansion rate decreases and when it becomes comparable to $m_{16}$, the scalar field $\Phi_{16}$ starts the coherent oscillation around its origin. 
After that, when $\Gamma(\Phi_{16}\rightarrow\psi_{8}+\psi_{8})\simeq H$ holds, $\Phi_{16}$ decays into the DMs.%
\footnote{Regarding the constraint from the isocurvature perturbations, the fluctuation of $\Phi_{16}$ is imprinted in the DMs in our mechanism.
 Thus, we assume $c_2\gtrsim \mathcal{O}(10)$ to suppress the isocurvature perturbations (see e.g. Ref.~\cite{Riotto:2002yw}).
Note that the fluctuation of the axial component of $\Phi_{16}$ is not suppressed by this way, but this does not matter because only the fluctuation of the radial component of $\Phi_{16}$ leads to the isocurvature perturbations of the DM.}
This mechanism is basically the same as the one discussed in Sec.~\ref{sec:fermionDMinflaton} whereas $\Phi_{16}$ production mechanism is different.

\begin{figure}[t]
\centering
\hspace*{-5mm}
\includegraphics[width=0.7\textwidth]{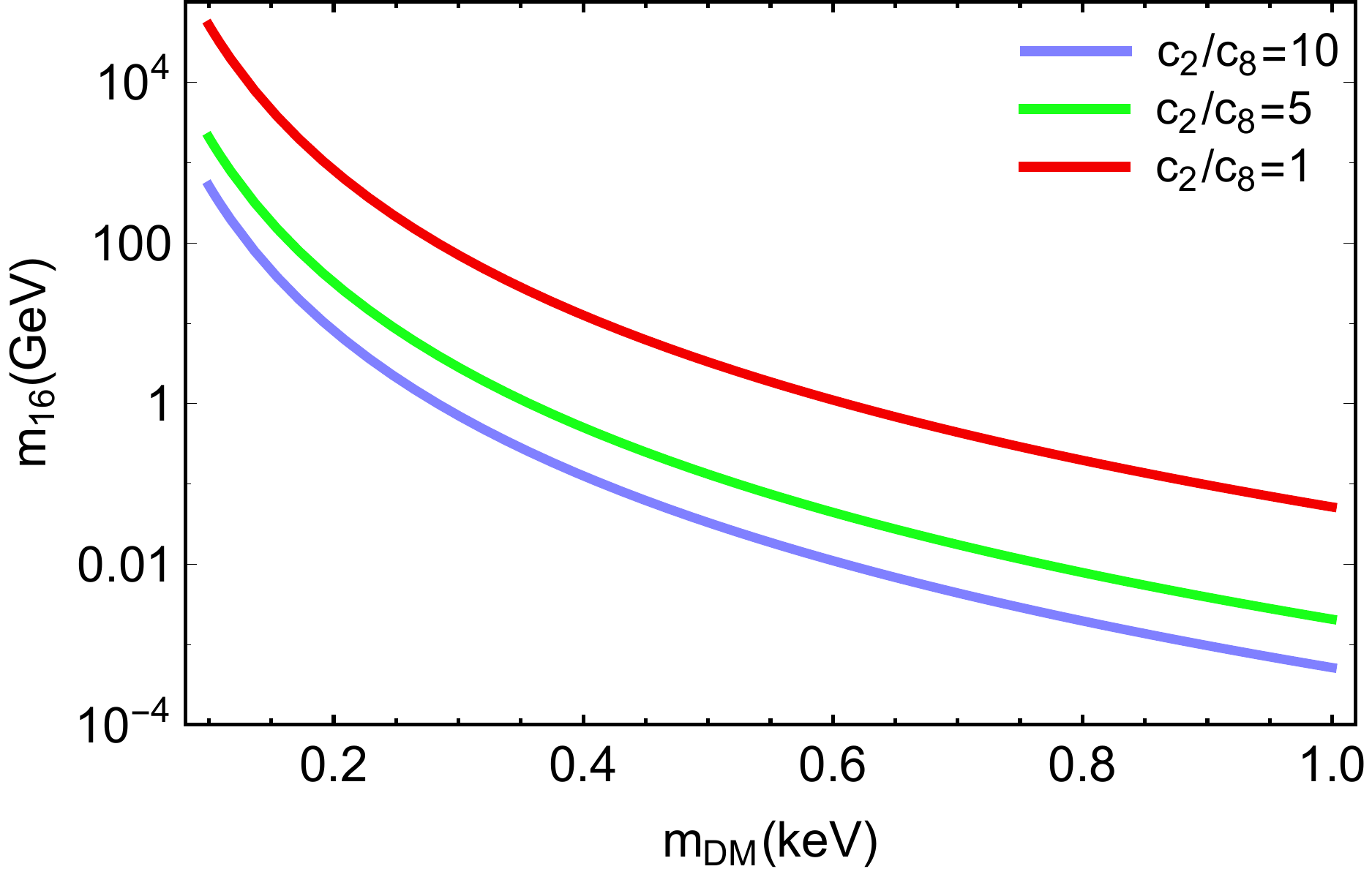}
\caption{The map between $m_{\rm DM}$ and $m_{16}$ obtained by DM relic density matching in Eq.~(\ref{eq:oscmatching2}).}
\vspace*{-1.5mm}
\label{f1}
\end{figure}

In the above DM production from the coherently oscillating $\Phi_{16}$, the abundance of DM is given by
\be
\left.\frac{2n_{\rm{16}}}{s_{\rm{SM}}}\right\vert_{a=a_{\rm osc}}\simeq\frac{m_{16}\Phi_{16,0}^{2}}{\frac{2\pi^{2}}{45}g_{*,s}(a_{\rm osc})T_{\rm osc}^{3}}\,,
\label{eq:oscmatching}
\ee
where $\Phi_{16,0}$ is the field amplitude of Eq.~(\ref{eq:Phi16amp}) when the oscillation of $\Phi_{16}$ starts ($H\simeq m_{16}$), $i.e.$
\be
\Phi_{16,0}\simeq  \left(\frac{(4!)^2c_2}{4\,c_{8}}m_{16}^2M_P^{4}\right)^{\frac{1}{6}}\ ,
\ee
$g_{*,s}$ is the effective degrees of freedom for the entropy density, and $T_{\rm osc}$ is the SM temperature at which the coherent oscillation of $\Phi_{16}$ occurs
\be
T_{\rm osc}=\left(\frac{90}{\pi^{2}}\right)^{1/4}g_{*}^{-1/4}(a_{\rm osc})\sqrt{M_{P}m_{16}}\simeq(0.85\times10^{9})\left(\frac{g_{*}(a_{\rm osc})}{100}\right)^{-1/4}\left(\frac{m_{16}}{1{\rm GeV}}\right)^{1/2}\,\,{\rm GeV}\,,
\label{eq:Tosc}
\ee
where $a_{\rm osc}$ is the scale factor as the oscillation starts.
Notice that we assumed that the oscillation starts at the radiation-dominated era $(T_{\rm osc}<T_R)$. By attributing the whole current DM abundance to $\psi_{8}$, we demand $2n_{16}/s_{\rm SM}=Y_{\rm DM}$ at $a=a_{\rm osc}$ which yields
\be
\left(\frac{g_{*}(a_{\rm osc})}{100}\right)^{-1/4}\left(\frac{m_{16}}{1{\rm GeV}}\right)^{1/6}\left(\frac{m_{\rm DM}}{1{\rm keV}}\right)\left(\frac{c_2}{c_{8}}\right)^{1/3}\simeq0.6\,.
\label{eq:oscmatching2}
\ee
This result tells us that for a given $c_2/c_8$
, there is one to one map between $m_{\rm DM}$ and $m_{16}$.
As an example, for $c_2/c_8=1,5,10$, we show this map in Fig.~\ref{f1}.

\begin{figure}[t]
\centering
\hspace*{-5mm}
\includegraphics[width=0.7\textwidth]{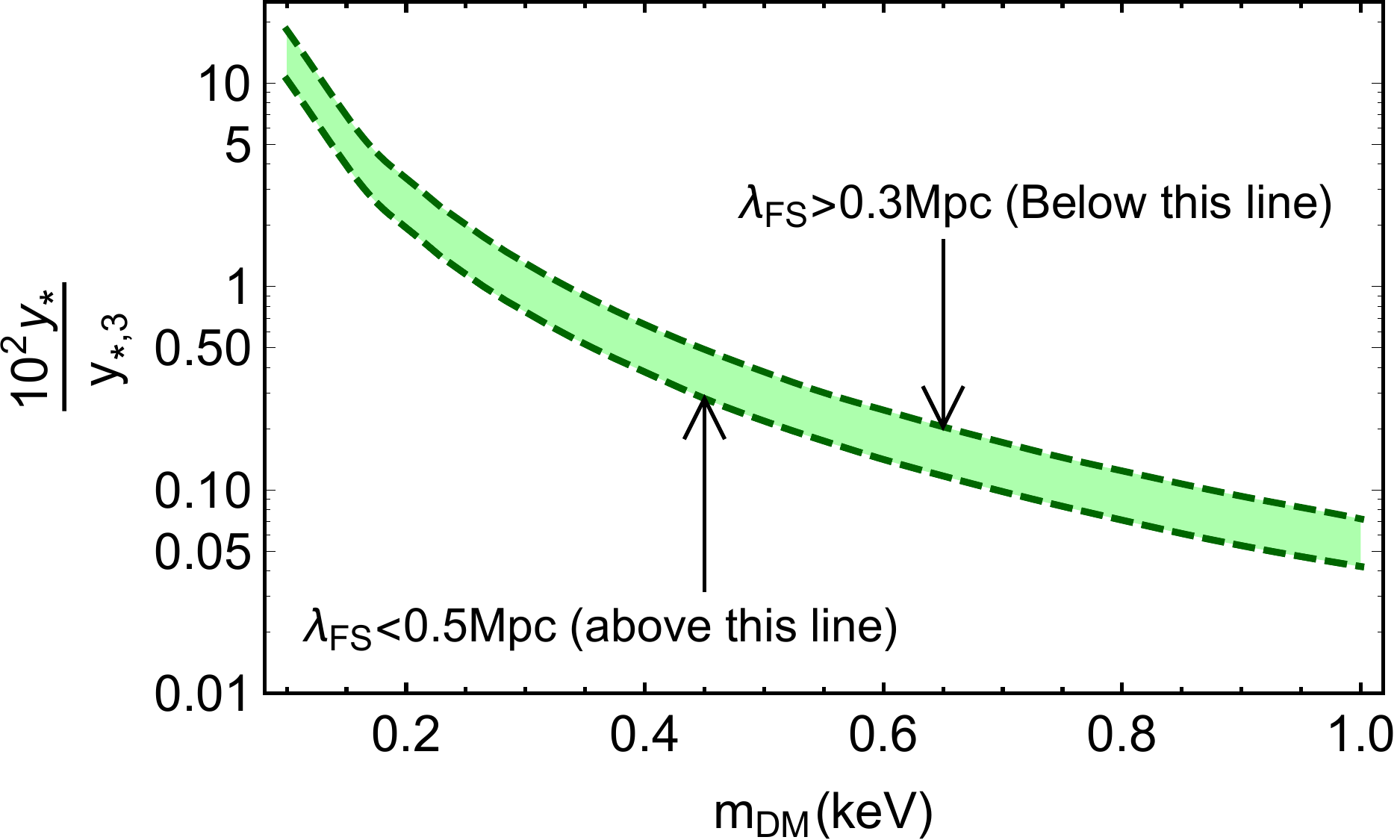}
\caption{For $c_{2}/c_{8}=5$ and $(m_{16},m_{\rm DM})$ given in Fig.~\ref{f1}, requiring $0.3{\rm Mpc}\!<\!\lambda_{\rm FS}\!<\!0.5$Mpc constrains the space of the Yukawa coupling between DM and $\Phi_{16}$.}
\vspace*{-1.5mm}
\label{f2}
\end{figure}

After the right amount of $\Phi_{16}$ is generated, in order to have $\psi_{8}$ as a degenerate fermion DM candidate today, we demand that
\be
T_{\rm osc}\,\,>\,\,T_{\rm SM,2}
\label{eq:hierarchy4}
\ee
where $T_{\rm SM,2}$ 
was defined in Eq.~(\ref{eq:T2}).
This leads to
\be
y_{*}\,\,<\,\,2.7\times\left(\frac{g_{*}(a_{\rm osc})}{100}\right)^{-1/4}\equiv y_{*,1}\,.
\label{eq:ystar31}
\ee
In addition, as discussed in Sec.~\ref{sec:nonzerolambda16}, we require
\begin{align}
y_*\,\,\lesssim\,\, (10^{-6})\left(\frac{m_{\rm DM}}{1{\rm keV}}\right)^{1/5} \left(\frac{m_{16}}{1{\rm GeV}}\right)^{3/10}\equiv y_{*,3}\ ,
\end{align}
so that the DMs do not form thermal bath via their self-interaction through $\Phi_{16}$ exchanges after its production. Define 
 $y_{*,{\rm max}}\equiv{\rm min}(y_{*,1},y_{*,3})$. 
Now for a set of $(m_{16}, m_{\rm DM}, c_{2}/c_{8})$ satisfying Eq.~(\ref{eq:oscmatching2}),  $y_{*,{\rm max}}$ is determined, and by choosing a $y_{*}\lesssim y_{*,{\rm max}}$, $\lambda_{\rm FS}$ can be computed based on Eq.~(\ref{eq:FS2}) and required to be $0.3{\rm Mpc}\!<\!\lambda_{\rm FS}\!<\!0.5$Mpc. For an example of $c_{2}/c_{8}=5$, we go through this procedure to constrain the space of the Yukawa coupling between DM and $\Phi_{16}$, of which the result is shown in Fig.~\ref{f2}. 

For this $y_{*}$, it turns out that $\Phi_{16}$ decay takes place before BBN era ($a_{p}\simeq\mathcal{O}(10^{-15})-\mathcal{O}(10^{-11}$)) and therefore sub-keV DM contributes to $\Delta N_{\rm eff}^{\rm BBN}$. Based on Eq.~(\ref{eq:Neff}), we compute $\Delta N_{\rm eff}^{\rm BBN}$ attributable to DM and find it is at most $\sim0.01$ to be consistent with $\Delta N_{\rm eff}^{\rm BBN}\lesssim0.114$ (95\% C.L.) \cite{Fields:2019pfx}.

From Eq.~(\ref{eq:wouldbeT}), we also estimate DM's ``would-be" temperature today for the case with $c_{2}/c_{8}=5$. The results read
$\tilde{T}_{{\rm DM,0}}\sim\mathcal{O}(10^{-8})-\mathcal{O}(10^{-7})$K which is smaller than $T_{{\rm DEG}}\simeq\mathcal{O}(10^{-4}){\rm K}-\mathcal{O}(10^{-3}){\rm K}$. This shows that current temperature of DM becomes low enough to accomplish the degenerate configuration when structure formation is ignored. 
We do not go further to discuss the cases with different ratios of $c_2/c_8$.
If one finds $\tilde{T}_{{\rm DM,0}}>T_{{\rm DEG}}$, one may arrive at a value of $c_{2}/c_{8}$ which is not allowed. But we note that $m_{16}$ and $a_{\rm FS}$ are inversely correlated in Eq.~(\ref{eq:wouldbeT}).

\section{Conclusion}
In this paper, we present a well-motivated extension of the SM which can address the core-cusp problem by providing a degenerate sub-keV fermion DM candidate. The model is characterized by $U(1)_{\rm B-L}$ gauge symmetry, and two right-handed heavy neutrinos and four new chiral fermions added to the SM gauge sector and particle contents respectively. All the fermions in the model are charged under $U(1)_{\rm B-L}$ and assigned the corresponding $Q_{\rm B-L}$s in a way that $U(1)_{\rm B-L}$ is rendered anomaly-free. It is extremely remarkable that one of the additional fermions obtains naturally a mass of $\mathcal{O}(1){\rm keV}$ because of its large B-L charge, provided that the B-L symmetry breaking scale $\sim10^{15}{\rm  GeV}$. Thus, it was shown that the chiral fermion can serve as a sub-keV fermion DM candidate of which temperature today is low enough to form a degenerate fermion halo core for a dSphs. The DM's free-streaming length is small enough to be consistent with Lyman-$\alpha$ forest data. Being WDM, the DM candidate in the model is also expected to resolve other small scale problems that $\Lambda$CDM paradigm confronts (the missing satellite and too-big-to-fail problem). Consequently, the model can resolve the small scale issues in cosmology as well as the smallness of the active neutrino mass and the baryon asymmetry via the thermal leptogenesis.

Concerning the DM production mechanism, we argue that fermion DM produced from the decay of a complex scalar can meet the criteria for a degenerate fermion DM. In Sec.~\ref{sec:SMscattering}, we showed that non-thermal DM produced from the SM particle scattering is bound to travel too large a free-streaming length. In Sec.~\ref{sec:fermionDMinflaton}, we showed that DM produced from a series of decays (inflaton decay and $\Phi_{16}$ decay) as the final product can travel the right size of the free-streaming length $\sim\mathcal{O}(0.1)$Mpc to be consistent with Lyman-$\alpha$ forest observation. Getting into more detail, we conducted the case study depending on whether a dark thermal bath forms (Sec.~\ref{sec:nonzerolambda16}) or not (Sec.~\ref{sec:zerolambda16}). For both cases, $\lambda_{\rm FS}$ for a fixed $m_{\rm DM}$ are parametrized by $(m_{16},y_{*})$. We figure out that for a set of $(m_{16},m_{\rm DM})$, the constraint applied to a choice of $y_{*}$ is more stringent for the case with formation of a dark thermal bath (Sec.~\ref{sec:nonzerolambda16}) than the other case (Sec.~\ref{sec:zerolambda16}). This fact makes it easier for the case without a dark thermal bath to produce a degenerate fermion DM consistent with the free-streaming length criterion. In Sec.~\ref{sec:fermionDMCO}, we studied a different mechanism to produce the degenerate fermion DM via the decay of a scalar field coherent oscillation. Differing from Sec.~\ref{sec:fermionDMinflaton} where a positive Hubble induced mass is assumed during inflation, a negative Hubble induced mass during inflation is assumed in Sec.~\ref{sec:fermionDMCO}. We studied a potential of $\Phi_{16}$ in Eq.~(\ref{eq:V16}) by which $\Phi_{16}$ field is located away from the origin in the field space at the end of the inflation. For a fixed $c_{2}/c_{8}$, there is one to one map between $m_{\rm DM}$ and $m_{16}$, which is required by DM relic density matching. Taking, for example, $c_{2}/c_{8}=5$, we showed that how the free-streaming length criterion $0.3{\rm Mpc}\lesssim\lambda_{\rm FS}\lesssim0.5{\rm Mpc}$ can constrain Yukawa coupling between the mother scalar field with $\sim m_{16}\in(10^{-3},10^{3}){\rm GeV}$ and DM candidate. For all distinct DM production mechanisms, we also performed further consistency checks including $\Delta N_{\rm eff}^{\rm BBN}$ contributed by DM and $\tilde{T}_{\rm DM,0}<T_{\rm DEG}$. Finally, we note that the framework presented in this paper shows that even if fermion warm DM mass is as low as sub-keV regime, it can still travel the free-streaming length as short as $\sim\mathcal{O}(0.1){\rm Mpc}$ consistent with Lyman-$\alpha$ forest observation thanks to the non-trivial dark sector structure and its cosmological history.

\section*{Acknowledgements}
T. T. Y. is supported in part by the China Grant for Talent Scientific Start-Up Project and the JSPS Grant-in-Aid for Scientific Research No. 16H02176, No. 17H02878, and No. 19H05810 and by World Premier International Research Center Initiative (WPI Initiative), MEXT, Japan. M. S. and T. T. Y. thank Kavli IPMU for their hospitality during the corona virus outbreak.
\newpage
\appendix

\vspace*{10mm}
\noindent
{\bf\LARGE Appendix}

\vspace*{1mm}
\section{$\xi$ decay}
\label{app:A}
$\xi$ is expected to decay to SM Higgs and a lepton via the decay operator
\be
\mathcal{O}_{\rm de}=\beta\frac{\left(\Phi_{-2}^*\right)^2}{M_P}\psi_{-5}\overline{N}\,
\label{eq:deoperator}
\ee
where $\beta$ is a dimensionless coefficient. 
For  $V_{\rm B-L}\sim3\times10^{15}$GeV, $m_\xi\simeq 2\times 10^9$\,GeV.
When the mass of the lightest right handed neutrino is about $10^9$\,GeV and its mass mixing with $\xi$ is $\mathcal{O}(1)$, $\xi$ can immediately decay into a Higgs and a lepton via the mixing once $\xi$ becomes non-relativistic.

\vspace*{1mm}
\section{Higgs Portal}
\label{app:B0}
\begin{figure}[h]
\centering
\vspace*{-20mm}
\includegraphics[width=0.8\textwidth]{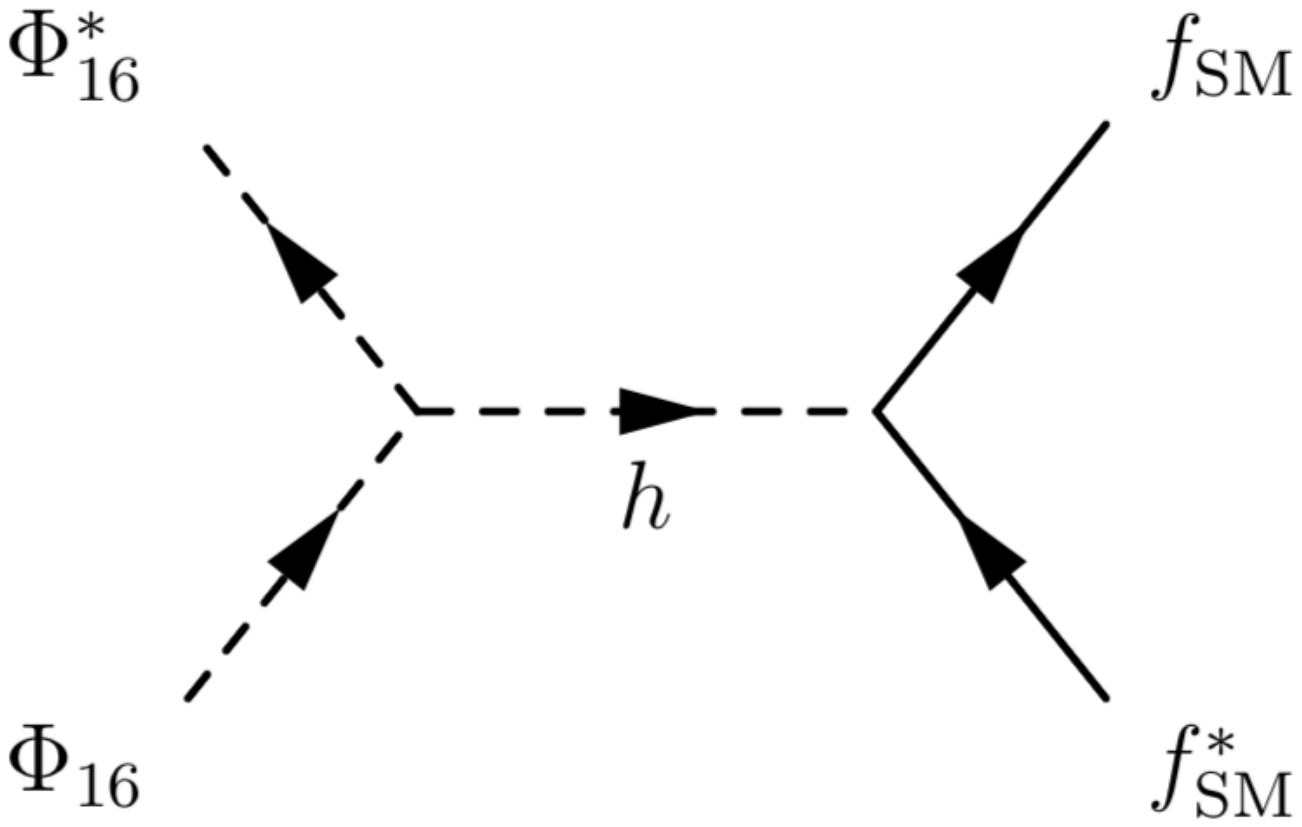}
\vspace*{-25mm}
\caption{For $m_{16}<\Lambda_{\rm EW}$, $\Phi_{16}$ is easily pair-annihilate to SM fermion pairs. $h$ is the Higgs field fluctuation around the global minimum of its potential.}
\vspace*{-1.5mm}
\label{fig4}
\end{figure}

The Higgs portal operator $\sim \lambda_{*}(H^{\dagger}H)|\Phi_{16}|^{2}$ allows for direct coupling between $\Phi_{16}$ and SM sector at the renormalizable level. The interaction rate for the process $H^{*}+H\rightarrow\Phi_{16}^{*}+\Phi_{16}$ owing to the Higgs portal operator, i.e. $\sim\lambda_{*}^{2}T$, is relatively much larger than not only interaction rate for the process $f_{\rm SM}^{*}+f_{\rm SM}\rightarrow\Phi_{16}^{*}+\Phi_{16}$ via $U(1)_{\rm B-L}$ gauge boson exchange, $\sim T^{5}/V_{\rm B-L}^{4}$, but also Hubble expansion rate since reheating time unless the Higgs portal is greatly suppressed. This tells us that produced from scattering among SM Higgs, $\Phi_{16}$ would be easily thermalized by SM thermal bath with significant $\lambda_{*}$. Once $\Phi_{16}$ joins the SM thermal bath,  trivially it never decouples. For the case where $\Phi_{16}$ decays before $\Phi_{16}$ becomes non-relativistic, $\psi_{8}$ becomes thermal WDM%
\footnote{The abundance of the WDM will be larger than the current dark matter abundance.}
 which is out of our interest. On the contrary, if $\Phi_{16}$ becomes non-relativistic before its decay to the DM starts, $\Phi_{16}$ would disappear prior to production of $\psi_{8}$.\footnote{If $\Phi_{16}$ is heavier than EW symmetry breaking scale, it will be Boltzmann suppressed once $T_{\rm SM}\simeq m_{16}$ is reached. If it is lighter than EW symmetry breaking scale, 
$\Phi_{16}$ is still living in the SM thermal bath by interaction with SM fermions induced by virtual SM Higgs. By comparing the relevant interaction rate of the diagram in Fig.~\ref{fig4} to Hubble expansion rate
\be
\Gamma\simeq\frac{\lambda_{*}^{2}m_{f}^{2}}{m_{h}^{4}}T_{\rm SM}^{3}\simeq\frac{T_{\rm SM}^{2}}{M_{P}}\simeq H\quad\Rightarrow\quad T_{\rm SM}\simeq\frac{m_{h}^{4}}{\lambda_{*}^{2}m_{f}^{2}M_{P}}
\ee
it is realized that $\Phi_{16}$ would easily pair-annihilate to SM fermions at $T_{\rm SM}\simeq m_{16}$. Here $m_{f}$ is a SM fermion mass and $m_{h}$ is the physical Higgs particle mass.} For these reasons, for the purpose of having sub-keV non-thermal fermion WDM, it is necessary for us to assume a highly suppressed Higgs portal operator $\sim \lambda_{*}(H^{\dagger}H)|\Phi_{16}|^{2}$.

\vspace*{1mm}
\section{$\Delta N_{\rm eff}$ contributed by DM ($\psi_{8}$)}
\label{app:B}
Recalling the expression for the radiation energy density
\be
\rho_{\rm rad}(T\lesssim1{\rm MeV})\simeq\rho_{\gamma}\left(1+\frac{7}{8}\left(\frac{4}{11}\right)^{4/3}N_{\rm eff}\right)\,,
\ee
we compute the extra-contribution to radiation from the relativistic DM at BBN era by
\be
\Delta N_{\rm eff}^{\rm BBN}\simeq\frac{\rho_{{\rm DM}}}{\rho_{\gamma}}\times\frac{8}{7}\left(\frac{11}{4}\right)^{4/3}\,,
\label{eq:Neff}
\ee
where based on Eq.~(\ref{eq:DMA1}), DM energy density at BBN time reads
\beq
\rho_{{\rm DM}}(a_{\rm BBN})&=&\sqrt{m_{{\rm DM}}^{2}+\left(\frac{m_{16}a_{\rm FS}}{2a_{\rm BBN}}\right)^{2}}\times\left[4.07\times10^{-4}\times\left(\frac{m_{{\rm DM}}}{1{\rm keV}}\right)^{-1}\right]\cr\cr
&&\times\frac{2\pi^{2}}{45}g_{s,{\rm SM}}(a_{{\rm BBN}})T_{{\rm SM}}(a_{\rm BBN})^{3}\,,
\eeq
and photon density is 
\be
\rho_{\gamma}(a_{\rm BBN})=\frac{\pi^{2}}{30}\times2\times(1{\rm MeV})^{4}
\ee

\vspace*{1mm}
\section{Mapping the thermal WDM mass to a non-thermal WDM}
\label{app:C}
It was observed in Ref.~\cite{Kamada:2013sh} that the linear matter power spectra associated with different WDM models are very similar when the same variance of velocity and the comoving Jean scale ($k_{J}$) are assumed. The comoving Jean scale at the matter-radiation equality time is defined as \cite{Kamada:2013sh} 
\be
\left.k_{J}=a\sqrt{\frac{4\pi G\rho_{m}}{\sigma^{2}}}\right\vert_{a=a_{\rm eq}}\,,
\label{eq:kJ}
\ee
where $\rho_{m}$ is the matter density and $\sigma$ is velocity variance of DM.

In accordance with this, it was argued in Ref.~\cite{Kamada:2019kpe} that equating the warmness parameters for the thermal WDM and WDM of another type differing from the thermal one constructs the map between masses. The warmness parameter ($\sigma\equiv\tilde{\sigma}T/m$) of a WDM introduced in \cite{Kamada:2019kpe} is defined with temperature $T$, mass $m$ and the quantity
\be
\tilde{\sigma}\equiv\frac{\int dq q^{4}f(q)}{\int q q^{2}f(q)}\,,
\label{eq:sigmatilde}
\ee
where $f(p)$ is the momentum space distribution function and $q\equiv p/m$ is used. To establish the map from the thermal WDM mass to another WDM candidate ($\chi$), one can begin with \begin{equation}
\sigma_{\chi}=\sigma_{\rm wdm} \Longleftrightarrow{} \tilde{\sigma}_{\chi}\frac{T_{_{\chi}}}{m_{\chi}}=\tilde{\sigma}_{\rm wdm}\frac{T_{\rm wdm}}{m_{\rm wdm}}\,,
\label{eq:Tmapping1}
\end{equation}
where $\sigma_{\chi}$ is the warmness of $\chi$-WDM and $\sigma_{\rm wdm}$ is that of the early decoupled thermal WDM. This equation tells us that once one knows $T_{\chi}$, $T_{\rm wdm}$ and $\tilde{\sigma}_{\chi}$ at $a=a_{\rm eq}$, one can map the constraint on $m_{\rm wdm}$ to that on $m_{\chi}$, knowing $\tilde{\sigma}_{\rm wdm}$=3.6 from Fermi-Dirac distribution. $T_{\chi}$ and $\tilde{\sigma}_{\chi}$ are closely related to production mechanism of $\chi$-WDM. On the other hand, for the early decoupled thermal WDM, $T_{\rm wdm}$ is determined by DM relic density. Today, comparison of thermal WDM to the neutrino gives \cite{Viel:2005qj} 
\begin{equation}
\Omega_{\rm wdm}h^{2}\simeq0.12=\left(\frac{m_{\rm wdm}}{94{\rm eV}}\right)\left(\frac{T_{\rm wdm,0}}{T_{\nu,0}}\right)^{3}\Longleftrightarrow{}T_{\rm wdm,0}=\left[0.036\left(\frac{94{\rm eV}}{m_{\rm wdm}}\right)\right]^{1/3}T_{\gamma,0}\,.,
\label{eq:Tmapping2}
\end{equation}
where $T_{\nu,0}=(4/11)^{1/3}T_{\gamma,0}$ is today's neutrino temperature.

\vspace*{1mm}
\section{Would-be temperature of DM candidate}
\label{app:D}
The necessary condition that fermion DM candidate should satisfy to form a cored halo profile within a dSphs is that its ``would-be" temperature today ($\tilde{T}_{{\rm DM,0}}$) in the absence of structure formation should be smaller than a degeneracy temperature for the dSphs ($T_{{\rm DEG}}$) \cite{Domcke:2014kla}. From the property that DM's momentum scales as $\sim a^{-1}$ and the temperature of DM can be defined via $E_{k}\sim kT$, we can infer that DM's temperature scales as $\sim a^{-1}$ for relativistic state and $\sim a^{-2}$ for non-relativistic state.

For the case where fermion DM candidate is produced from a non-relativistic scalar decay and free-stream since then, the scale factor ($a_{\rm NR}$) at which DM becomes non-relativistic is given by
\be
a_{\rm NR}\simeq\frac{m_{S}a_{\rm FS}}{2m_{\rm DM}}\,,
\label{eq:aNR}
\ee
where $m_{S}$ is the mother scalar's mass and $a_{\rm FS}$ is the scale factor at which DM starts free-streaming. Therefore, starting with $p_{\rm DM}(a_{\rm FS})\simeq m_{S}/2$, the ``would-be" temperature for DM today is computed by
\be
\tilde{T}_{\rm DM,0}\simeq\frac{m_{S}a_{\rm FS}}{2a_{\rm NR}}\times\left(\frac{a_{\rm NR}}{a_{0}}\right)^{2}=m_{\rm DM}\left(\frac{m_{S}a_{\rm FS}}{2m_{\rm DM}}\right)^{2}\,,
\label{eq:wouldbeT}
\ee
where Eq.~(\ref{eq:aNR}) is used for the second equality. The degeneracy temperature for a dSphs used for checking is roughly $T_{{\rm DEG}}\simeq\mathcal{O}(10^{-4}){\rm K}-\mathcal{O}(10^{-3}){\rm K}$ \cite{Domcke:2014kla}.

\newpage
\bibliographystyle{JHEP}
\bibliography{main}

\end{document}